\documentclass[]{iopart}
\usepackage{graphicx}
\usepackage{dcolumn}
\usepackage{amssymb}
\usepackage{ulem}
\usepackage{iopams}
\usepackage{bm}
\begin{document}
\newcommand{\be}{\begin{equation}}
\newcommand{\ee}{\end{equation}}
\newcommand{\ba}{\begin{eqnarray}}
\newcommand{\ea}{\end{eqnarray}}
\newcommand{\vk}{\mathbf{k}}
\newcommand{\vq}{\mathbf{q}}
\newcommand{\vp}{\mathbf{p}}
\newcommand{\vecr}{\mathbf{r}}
\newcommand{\vecR}{\mathbf{R}}
\newcommand{\vx}{\mathbf{x}}
\newcommand{\vy}{\mathbf{y}}
\newcommand{\vz}{\mathbf{z}}
\newcommand{\sfH}{\widehat{\mathsf{H}}}
\newcommand{\caH}{\widehat{\mathcal{H}}}
\newcommand{\la}{\langle}
\newcommand{\ra}{\rangle}
\newcommand{\half}{\frac{1}{2}}
\newcommand{\vv}{\mathbf{v}}
\newcommand{\tf}{t_{\mathsf{F}}}
\newcommand{\ti}{t_{\mathsf{I}}}
\newcommand{\U}{\widehat{\mathsf{U}}}
\title[Inelastic light scattering in Keldysh-Schwinger formalism]{
Study of resonant inelastic light scattering in Keldysh-Schwinger functional integral formalism}

\author{ H C Lee}
\address{ Department of Physics and Basic Science Research
Institute, Sogang University, Seoul, 121-742, Korea} 
\ead{hyunlee@sogang.ac.kr}
\begin{abstract}
The  scattering cross section of the resonant inelastic light scattering  is represented as  a correlation function 
in the Keldysh-Schwinger functional integral formalism.
The functional integral approach enables us to compute the cross section in the Feynman diagram perturbation theory where 
many-body effects can be fully incorporated. This approach is applied to the one G-phonon Raman scattering 
of graphene, and the result is shown to agree with the one previously obtained by the conventional Fermi golden rule formula.
Also, this approach is generalized to the systems in non-equilibrium conditions.
\end{abstract}
\submitto{\NJP}
\maketitle
\section{Introduction}
\label{intro}
The inelastic light scattering (ILS) is a very important experimental method 
for the investigation of the physical properties of condensed matter.
In particular, the resonant inelastic X-ray scattering (often called RIXS) has developed 
very rapidly recently, mainly owing to the advent of the high power X-ray light sources \cite{review}.
RIXS has proven to be a versatile tool for the study of elementary excitations of strongly correlated electron materials, such as 
magnons \cite{haverkort} and orbitons,
where many-body interactions play crucial roles \cite{review}.
Thus it is very desirable to have a theoretical scheme for the interpretation of the experimental 
ILS data which can incorporate many-body effects 
precisely.

In general, the transition rate $w_{\rm ILS}$ for ILS can be obtained by Fermi  golden rule
which is applied up to the second order with respect to the 
interaction Hamiltonian $\hat{\mathsf{H}}'$ between light and matter: 
\be
\label{golden}
w_{\rm ILS} = \frac{2\pi}{\hbar} \sum_{\mathsf{I}} \, p_{\mathsf{I}}\, 
\sum_{\mathsf{F}} \Big \vert \la \mathsf{ F} \vert \hat{\mathsf{H}}'_{\rm NR} \vert \mathsf{I} \ra 
+\sum_{\mathsf{n}}  \frac{\la \mathsf{F} \vert \sfH'_{\rm R} \vert \mathsf{n} \ra 
\la \mathsf{n} \vert \sfH'_{\rm R} \vert \mathsf{I} \ra }{E_{\mathsf{I}}-E_{\mathsf{n}}} \Big \vert^2,
\ee
where $ \vert \mathsf{I} \ra$, $\vert \mathsf{n} \ra$, and $\vert \mathsf{F} \ra$ denote the initial states, 
the (many-body) intermediate eigenstates, 
and the final states of the light-matter system, respectively. 
$E_{\mathsf{I},\mathsf{n}}$ are the energy eigenvalues of the states and $p_{\mathsf{I}}$ is the probability for the initial states 
($\sum_{\mathsf{I}} p_{\mathsf{I}} = 1 $).
$\hat{\mathsf{H}}'_{\rm NR}$ is the part of the Hamiltonian $\hat{\mathsf{H}}'$ \textit{quadratic}
in the photon vector potential operator $\vec{A}$, 
which contributes to the \textit{nonresonant} ILS \cite{shastry}. 
$\hat{\mathsf{H}}'_{\rm R}$ is \textit{linear} in $\vec{A}$, and it is responsible for the \textit{resonant} ILS \cite{shastry}
[the second term of (\ref{golden})].
The nonresonant transition amplitude  often dominates over the resonant one,
but in special cases where $E_{\mathsf{I}}$ is close to $E_{\mathsf{n}}$, the resonant transition amplitude can be significant.
After some manipulations involving the expectation value of the photon operators, (\ref{golden}) 
can be shown to be essentially the Kramers-Heisenberg formula of ILS.
In this paper, for simplicity, we will consider only the case 
$
\hat{\mathsf{H}}'_{\rm R} = -\frac{1}{c} \int d \vec{r} \,  \vec{A}(\vec{r}) \cdot \vec{J}_{\rm e}(\vec{r}),
$
where $ \vec{J}_{\rm e}(\vec{r})$ is the electric current operator.

It is well known that the scattering cross-section of nonresonant contribution can be expressed 
in the form of correlation function of stress tensor operators \cite{shastry,agd}
by using van Hove-Placzek procedure \cite{hove,placzek}.
By  employing the fluctuation-dissipation theorem the correlation function can be directly linked to
 the retarded Green's function (\textit{e.g.}, see  p.151 of \cite{agd}).
Then the retarded Green's function can be computed systematically (including full many-body effects)
in imaginary time Matsubara formalism \cite{agd,mahan}
by using analytic continuation technique.
Thus as far as the nonresonant contribution to ILS is concerned we have a firmly established theoretical scheme where full many-body 
physics can be incorporated precisely.
Henceforth we will assume that resonance condition is satisfied and will focus on the resonant contributions only.

The \textit{first} goal of this paper is to express the resonant contribution to ILS in the form of correlation function
just like the case of the nonresonant contribution mentioned above.
To achieve this goal, in principle,  we have to carry out the sum over the intermediate states $\vert \mathsf{n} \ra$ of
the second term of  (\ref{golden}). 
In general, very little is known about the states $\vert \mathsf{n} \ra$, and we have to employ
some kinds of assumptions or approximations about the states.
Correlation functions are defined without reference to intermediate states, so in our approach
the explicit sum over the intermediate states $\vert \mathsf{n} \ra$ should be avoided.
Then we have to go back to the time-dependent perturbation theory involving time evolution operator. 
Recall that the time-dependent perturbation theory  requires time-ordering of operators \cite{sakurai}, namely 
\be
\label{ordering}
\mathtt{T} ( \hat{X}(t) \hat{Y}(t') ) = \theta(t-t') \hat{X}(t) \hat{Y}(t') \pm \theta(t'-t) \hat{Y}(t') \hat{X}(t),
\ee
where minus sign applies when both $\hat{X}(t)$ and $\hat{Y}(t')$ are fermion operators and $\mathtt{T}$ is the time-ordering symbol
\cite{comment1}. $\theta(t)$ is the unit step function.

Now as can be seen in (\ref{golden}), we have to take the complex conjugate of transition amplitude to obtain the transition
rate. In quantum mechanics the complex conjugate of a  transition amplitude of a process  is related to
a transition amplitude of the \textit{time-reversed} process, 
and  this observation necessitates  the introduction of  \textit{anti}-time-ordering:
\be
\label{antiordering}
\tilde{\mathtt{T}} ( \hat{X}(t) \hat{Y}(t') ) = \theta(t'-t) \hat{X}(t) \hat{Y}(t') \pm \theta(t-t') \hat{Y}(t') \hat{X}(t),
\ee
where $\tilde{\mathtt{T}}$ is the anti-time-ordering symbol.

From the above discussion it is evident that the scattering cross section of ILS
(generically all transition probability) requires 
not only forward time evolution but also \textit{backward} time evolution 
in the \textit{time}  representation of 
the time-dependent perturbation theory. Therefore, we are naturally led to 
the concept of \textit{closed} time contour introduced by  Keldysh and Schwinger (KS) \cite{kel,schwinger}.

The KS formulation of time-dependent perturbation theory has been employed in the earlier works on 
the X-ray spectra of metals\cite{canad,doniach,nozieres}.
However, these works did not fully disentangle the  time orderings of involved operators, 
so that they were limited to the   time-representation of Feynman
diagrams or the time-ordered Feynman diagrams 
[see, for example,  the figure 2 of \cite{doniach} and the equation (11) of \cite{nozieres}]. 
If the time orderings are fully disentangled then all time-integrals can be done from $-\infty$ to $\infty$, and 
all Feynman diagrams can be calculated in energy-momentum representation.

The \textit{second} goal of this paper is to elaborate on the formalism developed in the references  \cite{canad,doniach,nozieres},
so that the correlation function obtained in the first goal can be recast in the KS functional integral formalism, 
which will disentangle the time orderings completely.
The correlation functions in KS functional integral formalism can be computed in the standard Feynman diagram method of 
quantum field theory \cite{peskin} without being limited to  time-ordered diagrams, etc.

The \textit{third} goal of this paper is to generalize the KS functional integral approach to the systems in non-equilibrium
conditions. The KS methods were originally developed for the study of non-equilibrium phenomena.
As such, the KS functional integral formalism can be readily generalized for the study of ILS in non-equilibrium situation.

The main results of this paper are (\ref{correlation}),  (\ref{cross5}), (\ref{main3}), (\ref{main4}), and (\ref{action-noneq}).
This paper is organized as follows.
In section \ref{scatter} we derive the correlation function corresponding to  the scattering cross-section of the resonant ILS.
In section \ref{functional} the correlation function  is represented in KS functional integral
formalism. The key step is the disentanglement of the time orderings of electric current operators.
In section \ref{comparison} we compare our approach with the Kramers-Heisenberg formula and point out differences.
In section \ref{application} we apply our method to the case of G-phonon Raman scattering of
graphene and compare the results  with the one obtained by conventional Fermi golden rule method.
In section \ref{noneq} we generalize the KS functional integral formalism of ILS to the non-equilibrium situations.
We close this paper with discussions and summary in section \ref{summary}.

\section{Scattering cross-section of ILS and correlation function}
\label{scatter}
First we recall that 
the  scattering cross-section  of ILS is defined by the transition rate (induced by  light-matter interaction)
divided by the flux of incoming photons: 
\be
\label{cross}
d \sigma = \sum_{\mathsf{I}} p_{\mathsf{I}}    \sum_{\mathsf{F}} 
 \frac{| \mathcal{ M}_{\mathsf{FI}}(t_{\mathsf{F}},t_{\mathsf{I}}) |^2}{(t_{\mathsf{F}}-t_{\mathsf{I}})}/ (\frac{c}{\mathcal{V}}),
\ee
where $\mathcal{ M}_{\mathsf{FI}}$ is the transition amplitude from an initial state $\vert \mathsf{I} \ra$ to a
final state $\vert \mathsf{F} \ra$ in time interval $[t_{\mathsf{I}}, t_{\mathsf{F}}]$
and $\mathcal{V}$ is the volume of space for the box quantization of photons.
Our goal in this section is to express (\ref{cross}) in the form of correlation function.
The key quantity is the transition amplitude $\mathcal{M}_{\mathsf{FI}}(\tf,\ti)$ which is most naturally
presented in the framework of  time-dependent perturbation theory \cite{sakurai}.

The transition amplitude $\mathcal{M}_{\mathsf{FI}}(\tf,\ti)$  is defined by 
\be
\label{amplitude}
\mathcal{M}_{\mathsf{FI}}(t_{\mathsf{F}},t_{\mathsf{I}}) =
\la \mathsf{F} \vert \widehat{\mathsf{U}}_{\mathsf{H}_{\rm tot}}(t_{\mathsf{F}},t_{\mathsf{I}}) \vert \mathsf{I} \ra, 
\ee
where $ \widehat{\mathsf{U}}_{\widehat{\mathsf{H}}_{\rm tot}}(t_{\mathsf{F}},t_{\mathsf{I}})$ is 
the time evolution operator governed by the 
total Hamiltonian ${\widehat{\mathsf{H}}_{\rm tot}}$ of light-matter system from time $t_{\mathsf{I}}$ to $t_{\mathsf{F}}$:
\be
\label{evolution}
 \widehat{\mathsf{U}}_{\widehat{\mathsf{H}}_{\rm tot}}(t_{\mathsf{F}},t_{\mathsf{I}})  =\mathtt{T} 
 \exp\Big[-\frac{i}{\hbar} \int^{t_{\mathsf{F}}}_{t_{\mathsf{I}}} \,  d \bar{t} \, \,\widehat{\mathsf{H}}_{\rm tot}(\bar{t})\Big ].
\ee
The total Hamiltonian consists of the Hamiltonian of matter $\sfH_{\rm m}$, 
the Hamiltonian of photons  $ \sfH_{\rm p}$ [see (\ref{photonH})] ,
and the interaction Hamiltonian between light and matter $\widehat{\mathsf{H}}'$:
\be
\label{total}
 \sfH_{\rm tot} = (\sfH_{\rm m}+  \sfH_{\rm p}) + \hat{\mathsf{H}}'
  \equiv \sfH_0 +\hat{\mathsf{H}}'.
  \ee

Next we employ the interaction picture  and treat $\widehat{\mathsf{H}}'$ perturbatively.
In the interaction picture the time evolution operator 
$\widehat{\mathsf{U}}_{\widehat{\mathsf{H}}_{\rm tot}}(t_{\mathsf{F}},t_{\mathsf{I}}) $
becomes (see, for example,  pp.722-724 of  \cite{messiah})
\be
\label{disen}
 \widehat{\mathsf{U}}_{\widehat{\mathsf{H}}_{\rm tot}}(t_{\mathsf{F}},t_{\mathsf{I}}) =
 \widehat{\mathsf{U}}_{\widehat{\mathsf{H}}_0}(t_{\mathsf{F}},t_{\mathsf{I}}) \,\widehat{\mathsf{V}}(t_{\mathsf{F}},t_{\mathsf{I}}),
 \ee
 where $\widehat{\mathsf{U}}_{\hat{\mathsf{H}}_0}(t_{\mathsf{F}},t_{\mathsf{I}})$ is the time evolution operator governed by the 
 Hamiltonian $\sfH_0$ [ see (\ref{total})], and 
 \be
 \label{interaction}
 \widehat{\mathsf{V}}(t_{\mathsf{F}},t_{\mathsf{I}}) = \mathtt{T} \exp \big[-\frac{i}{\hbar} \int^{t_{\mathsf{F}}}_{t_{\mathsf{I}}} \, d \bar{t} \,\,
 \sfH_{\sfH_0}^{\prime}(\bar{t}) \big ],
 \ee
 where $\sfH_{\sfH_0}^{\prime}$ denotes the Hamiltonian $\sfH'$
 in the Heisenberg picture with respect to $\sfH_{0}$.
 Now plugging (\ref{disen}) into  (\ref{amplitude}) and expanding 
 (\ref{interaction}) up to the second order in $\sfH_{\sfH_0}^{\prime}$, 
 we obtain  the second order contribution which is responsible for the resonant ILS:
\be
\label{secondM}
\mathcal{M}^{(2)}_{\mathsf{FI}} = \frac{1}{2!}\left( -\frac{i}{\hbar} \right)^2 
\la \mathsf{F} \vert \widehat{\mathsf{U}}_{\sfH_0} \, \int^{\tf}_{\ti} d \bar{t}_1  \int_{\ti}^{\tf} d \bar{t}_2 
\, \mathtt{T} [\sfH_{\sfH_0}'(\bar{t}_1) \sfH_{\sfH_0}'(\bar{t}_2)] \vert \mathsf{I} \ra.
\ee
Then the transition \textit{probability} can be expressed as
\ba
\label{prob}
\vert \mathcal{M}^{(2)}_{\mathsf{FI}} \vert^2 &=& \frac{1}{(2!)^2 \hbar^4}\la \mathsf{I} \vert  
 \int^{\tf}_{\ti} d \bar{t}_3  \int_{\ti}^{\tf} d \bar{t}_4\, \tilde{\mathtt{T}} 
[\sfH_{\sfH_0}'(\bar{t}_4) \sfH_{\sfH_0}'(\bar{t}_3)]  \widehat{\mathsf{U}}^\dag_{\sfH_0}\vert \mathsf{F} \ra \\
&\times &\la \mathsf{F} \vert \widehat{\mathsf{U}}_{\sfH_0} \, \int^{\tf}_{\ti} d \bar{t}_1  \int_{\ti}^{\tf} d \bar{t}_2 
\, \mathtt{T} [\sfH_{\sfH_0}'(\bar{t}_1) \sfH_{\sfH_0}'(\bar{t}_2)] \vert \mathsf{I} \ra.
\nonumber
\ea
Note the appearance of the anti-time ordering $\tilde{\mathtt{T}}$ in (\ref{prob}).

The next key step  is to compute the matrix elements of  (\ref{prob}) with respect to  the \textit{photon} states (only).
For that purpose,  
the initial  $\vert \mathsf{I} \ra$ and the final   $\vert \mathsf{F} \ra$ states are represented by 
the tensor product of  photon Hilbert space 
$ | \vk_{\mathrm{i/f}}, \hat{\epsilon}_{\mathrm{i/f}} \ra$  and  matter Hilbert space $ \vert \mathrm{i/f} \ra$:
\be
| \mathsf{I} \ra = | \vk_{\mathrm{i}}, \hat{\epsilon}_{\mathrm{i}} \ra \otimes |\mathrm{ i }\ra, \;
| \mathsf{F}  \ra = | \vk_{\mathrm{f}}, \hat{\epsilon}_{\mathrm{f}} \ra \otimes | {\rm f}  \ra,
\ee
where $\vk_{\rm i,f}$ and $\hat{\epsilon}_{\rm i,f}$ are the wavenumber vector and the polarization vector of incoming and 
outgoing photon, respectively.
Since
the matter Hamiltonian $\sfH_{\rm m}$ commutes with  the photon Hamiltonian $\sfH_{\rm p}$ which is non-interacting 
[see (\ref{photonH})], 
the action of $\sfH_{\rm p}$  and its time evolution operator $\mathsf{U}_{\sfH_{\rm p}}$  in (\ref{prob}) 
on photon Hilbert space can be evaluated exactly.
After evaluating the action of  $\mathsf{U}_{\sfH_{\rm p}}$, we can define
 an operator $\hat{\mathsf{D}}_{\tilde{k}_{\rm i},\tilde{k}_{\rm f}}$ and its Hermitian conjugate 
$\hat{\mathsf{D}}^\dag_{\tilde{k}_{\rm i},\tilde{k}_{\rm f}}$ which act \textit{ only on the matter Hilbert space}. 
\ba
\label{d1}
\widehat{\mathsf{D}}_{\tilde{k}_{\rm i},\tilde{k}_{\rm f}}(t_1,t_2) &=& \la \vk_{\rm f}, \hat{\epsilon}_{\rm f}\vert 
 \mathtt{T} [\sfH_{\sfH_0}'(\bar{t}_1) \sfH_{\sfH_0}'(\bar{t}_2)] \vert \vk_{\rm i},\hat{\epsilon}_{\rm i} \ra,  \\
\widehat{\mathsf{D}}^\dag_{\tilde{k}_{\rm i},\tilde{k}_{\rm f}}(t_3,t_4) &=& \la  \vk_{\rm i},\hat{\epsilon}_{\rm i} \vert \tilde{\mathtt{T}} 
[\sfH_{\sfH_0}'(\bar{t}_4) \sfH_{\sfH_0}'(\bar{t}_3)]  \vert  \vk_{\rm f}, \hat{\epsilon}_{\rm f} \ra,
\label{d2}
\ea
where $\tilde{k}_{\rm i,f}$ denotes  $\vk_{\rm i,f}$ and $ \hat{\epsilon}_{\rm i,f}$ collectively.

Then the transition probability (\ref{prob}) can be recast as
\be
\label{prob2}
\fl
\vert \mathcal{M}^{(2)}_{\mathsf{FI}} \vert^2 = \frac{1}{(2!)^2 \hbar^4} \,\prod_{l=1}^4 \int_{\ti}^{\tf} d \bar{t}_l
\la {\rm i} \vert \widehat{\mathsf{D}}^\dag_{\tilde{k}_{\rm i},\tilde{k}_{\rm f}}(\bar{t}_3,\bar{t}_4) 
\widehat{\mathsf{U}}_{\sfH_{\rm m}}^\dag \vert {\rm f} \ra \la {\rm f} \vert \widehat{\mathsf{U}}_{\sfH_{\rm m}} 
\widehat{\mathsf{D}}_{\tilde{k}_{\rm i},\tilde{k}_{\rm f}}(\bar{t}_1,\bar{t}_2) \vert {\rm i} \ra.
\ee
We note that the operator $\widehat{\mathsf{D}}_{\tilde{k}_{\rm i},\tilde{k}_{\rm f}}(t_1,t_2)$ is time-ordered, while
the operator $\widehat{\mathsf{D}}^\dag_{\tilde{k}_{\rm i},\tilde{k}_{\rm f}}(t_3,t_4)$ is \textit{anti}-time-ordered.

The initial photon state  $\vert \vk_{\rm i}, \hat{\epsilon}_{\rm i} \ra$  can be taken to a  pure state, so that the 
probability distribution for the initial state $p_{\mathsf{I}}$ pertains only to the initial \textit{matter} Hilbert space.
The probability distribution $p_{\rm i}$ for the initial matter Hilbert space can be specified by a density matrix $\hat{\rho}_{\rm m}$
( $\mathrm{Tr}_{\rm m}$ indicates trace over 
matter Hilbert space):
\be
\label{densityM}
p_{\rm i} = \la {\rm i} \vert \hat{\rho}_{\rm m} \vert {\rm  i} \ra, \quad  
\mathrm{Tr}_{\rm m} \hat{\rho}_{\rm m} =1.
\ee
The scattering cross-section of resonant ILS now takes the form 
\be
\label{cross2}
d \sigma =\frac{1}{c/\mathcal{V}}\,\frac{1}{\tf-\ti} \, \, \sum_{\vk_{\rm f},\hat{\epsilon}_{\rm f}} \, 
\sum_{\rm i}   \la {\rm i} \vert \hat{\rho}_{\rm m} \vert {\rm  i} \ra  \sum_{\rm f} \vert \mathcal{M}^{(2)}_{\mathsf{FI}} \vert^2.
\ee
The sum over the \textit{final} matter Hilbert space of (\ref{cross2}) can be done using the closure relation  of matter Hilbert space
$ \sum_{\rm f} \vert {\rm f} \ra \la {\rm f} \vert = {\rm I}$ ( I is an identity operator).
From (\ref{prob2}) we obtain
\be
\label{cross2-1}
\sum_{\rm f} \vert \mathcal{M}^{(2)}_{\mathsf{FI}} \vert^2 = 
 \frac{1}{(2!)^2 \hbar^4} \,\prod_{l=1}^4 \int_{\ti}^{\tf} d \bar{t}_l
\la {\rm i} \vert \widehat{\mathsf{D}}^\dag_{\tilde{k}_{\rm i},\tilde{k}_{\rm f}}(\bar{t}_3,\bar{t}_4)  
\widehat{\mathsf{D}}_{\tilde{k}_{\rm i},\tilde{k}_{\rm f}}(\bar{t}_1,\bar{t}_2) \vert {\rm  i} \ra.
\ee
The sum over the initial matter Hilbert space $\vert {\rm i} \ra$ of (\ref{cross2})  can done by using  (\ref{densityM}).
\be
\label{cross3}
\fl d \sigma = \frac{1}{(2!)^2 \hbar^4}\frac{1}{c/\mathcal{V}}\frac{1}{\tf-\ti} \,  \sum_{\vk_{\rm f},\hat{\epsilon}_{\rm f}} \,
\prod_{l=1}^4 \int_{\ti}^{\tf} d \bar{t}_l \,  \mathrm{Tr}_{\rm m} \Big[ \hat{\rho}_{\rm m} 
  \widehat{\mathsf{D}}^\dag_{\tilde{k}_{\rm i},\tilde{k}_{\rm f}}(\bar{t}_3,\bar{t}_4) 
  \widehat{\mathsf{D}}_{\tilde{k}_{\rm i},\tilde{k}_{\rm f}}(\bar{t}_1,\bar{t}_2)\Big ].
\ee
Equation (\ref{cross3}) represents the scattering cross section in the form of  correlation function 
of operators $\widehat{\mathsf{D}}_{\tilde{k}_{\rm i},\tilde{k}_{\rm f}}$ and 
$\widehat{\mathsf{D}}_{\tilde{k}_{\rm i},\tilde{k}_{\rm f}}^\dag$
which act only on the matter Hilbert space.
The detailed form of the operator $\widehat{\mathsf{D}}_{\tilde{k}_{\rm i},\tilde{k}_{\rm f}}$ is worked out 
in Appendix A. 

Substituting (\ref{op1}) and (\ref{op2}) into (\ref{cross3}) we arrive at
\be
\label{cross4}
\fl
d \sigma = \frac{1}{\tf-\ti} \frac{(2\pi/\hbar)^2}{c  \omega_{\rm i} \omega_{\rm f}} 
\, \sum_{\vk_{\rm f},\hat{\epsilon}_{\rm f}} \int (d t_i) \, e^{-i \omega_{\rm i} t_1 + i \omega_{\rm f} t_2 + i \omega_{\rm i} t_3 - i\omega_{\rm f} t_4}
C_{k_{\rm i} \hat{\epsilon}_{\rm i}, k_{\rm f} \hat{\epsilon}_{\rm f}}(t_3,t_4,t_1,t_2),
\ee
where the correlation function $C_{k_{\rm i} \hat{\epsilon}_{\rm i}, k_{\rm f} \hat{\epsilon}_{\rm f}}(t_3,t_4,t_1,t_2)$ 
is given by [$\vec{J}_{\rm e}(\vk, t)$ is the electric current operator in momentum space, see (\ref{fourier}) ]
\be
\label{correlation}
\fl
\eqalign{
C_{k_{\rm i} \hat{\epsilon}_{\rm i}, k_{\rm f} \hat{\epsilon}_{\rm f}}(t_3,t_4,t_1,t_2)
&=\mathrm{Tr}_{\rm m} \Big[ \hat{\rho}_{\rm m} 
\tilde{\mathtt{T}} [ \vec{J}_{\rm e}(\vk_{\rm i}, t_3) \cdot \hat{\epsilon}_{\rm i}^*  \, \,
\vec{J}_{\rm e}(-\vk_{\rm f}, t_4) \cdot \hat{\epsilon}_{\rm f}  ] \cr
&\times \mathtt{T} [ \vec{J}_{\rm e}(-\vk_{\rm i}, t_1) \cdot \hat{\epsilon}_{\rm i} \,  \,\vec{J}_{\rm e}(\vk_{\rm f}, t_2) \cdot 
\hat{\epsilon}^*_{\rm f} ] \Big ].
}
\ee
For matter systems at equilibrium  the correlation function $C_{k_i,k_f} (t_3,t_4,t_1,t_2)$ possesses
the time translation invariance which is manifest in the following form:
\be
\label{fourier2}
\fl
C_{k_{\rm i} \hat{\epsilon}_{\rm i}, k_{\rm f} \hat{\epsilon}_{\rm f}}(\{ t_i \}) 
= \int \prod_{j=1}^4 \frac{d \omega_j}{2\pi} e^{-i \sum_j \omega_j t_j} 
(2\pi) \delta( \sum_j \omega_j) \,\, \tilde{C}_{k_{\rm i} \hat{\epsilon}_{\rm i}, k_{\rm f} \hat{\epsilon}_{\rm f}}(\{ \omega_i \}).
\ee
Plugging (\ref{fourier2}) into (\ref{cross4}), 
changing variable $t_i  \to t_i + (\tf+\ti)/2$, and carrying out the time integrals 
in the limit $\tf-\ti \to \infty$, we obtain ($ \sum_{\vk_{\rm f}} = \mathcal{V} \, \int d^3 \vk_{\rm f}/(2\pi)^3$)
\be
\label{cross5}
 d \sigma =  \frac{(2\pi/\hbar)^2}{c \omega_{\rm i} \omega_{\rm f}} \,\int \frac{ d^3 \vk_{\rm f}}{(2\pi)^3}
\sum_{\hat{\epsilon}_{\rm f}} 
\tilde{C}_{k_{\rm i} \hat{\epsilon}_{\rm i}, k_{\rm f} \hat{\epsilon}_{\rm f}}(\omega_{\rm i},-\omega_{\rm f}, -\omega_{\rm i},
\omega_{\rm f} ),
\ee
where  the energy conserving delta function is substituted by
\be
2 \pi \delta (0) = \int^{(\tf-\ti)/2}_{-(\tf-\ti)/2} dt e^{i t \cdot 0} \to (\tf-\ti).
\ee
(\ref{correlation}) and (\ref{cross5}) constitute the first goal of this paper.

\section{Keldysh-Schwinger functional integral formulation of correlation function}
\label{functional}
As we have seen in section \ref{scatter}
the computation of scattering cross-section has been reduced to the computation of the correlation function 
(\ref{correlation}), which we rephrase as
\be
\label{cor1}
\Pi(\tau_1,\tau_2,\tau_3,\tau_4)=\mathrm{Tr}_{\rm m} \Big( \,\,\tilde{\mathtt{T}} [\widehat{A} (\tau_1) \widehat{B} (\tau_2) ] \,
                            \mathtt{T} [\widehat{C} (\tau_3) \widehat{D}(\tau_4) ] \, \hat{\rho}_{\rm m} \Big )
\ee
for the sake of more convenient notations. From now on we will  drop the subscript  ``m'' denoting the matter sector.
In this section the correlation function (\ref{cor1}) will be  represented in  KS functional integral formulation, 
and in this representation   all time-orderings will be  disentangled completely.

The literature on the method of KS closed time contour are very vast, and we cite only 
\cite{kamenev,rammer} which appear to be more useful for condensed matter physicists in author's opinion.
We will also need the method of coherent state path integral, and the detailed treatments can be found in \cite{orland}.
In the following many well-known detailed steps (for which readers are referred to the references cited above)
will be omitted and we will focus on the disentangling the time orderings of 
(\ref{cor1}). 

\begin{figure}[!hbt]
\begin{center}
\includegraphics[width = 0.5 \textwidth]{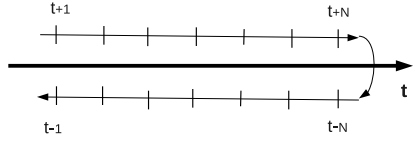}
\caption{The closed KS time contour is depicted. 
The upper branch is for the forward time evolution $\widehat{\mathsf{U}}(\tf,\ti)$, while the lower branch is 
for the backward time evolution $\widehat{\mathsf{U}}(\ti,\tf)$.}
\label{contour}
\end{center}
\end{figure}

The key concept  of the KS method is the time evolution along  \textit{closed} time contour.
A characteristic property of this closed time contour $\mathcal{C}=[\ti\to\tf \to \ti]$  (see figure \ref{contour}) is 
(where we put $\hat{\rho}_{\rm m} = \hat{\rho}(\ti)$)
\be
\label{partition}
Z  \equiv \mathrm{Tr} \Big[ \widehat{\mathsf{U}}(\ti, \tf) \widehat{\mathsf{U}}(\tf,\ti) \hat{\rho}(\ti) \Big ] =
\mathrm{Tr} \Big[ \widehat{\mathsf{U}}_{\mathcal{C}}\hat{\rho}(\ti) \Big ] =
1,
\ee
which  follows from the unitarity of time evolution operator $\widehat{\mathsf{U}}$.
We start by resolving the time orderings of (\ref{cor1}) explicitly:
\ba
\label{ordering1}
\tilde{\mathtt{T}} [ \widehat{A}(\tau_1) \hat{B} (\tau_2) ] &=& \theta(\tau_2 -\tau_1)  \widehat{A}(\tau_1) \hat{B} (\tau_2) +
\theta(\tau_1-\tau_2)  \widehat{B} (\tau_2) \widehat{A}(\tau_1)  \cr
\mathtt{T} [ \widehat{C}(\tau_3) \widehat{D} (\tau_4) ] &=& \theta(\tau_3 -\tau_4)  \widehat{C}(\tau_3) \widehat{D} (\tau_4) +
\theta(\tau_4-\tau_3)  \widehat{D} (\tau_4) \widehat{C}(\tau_3).
\ea
Plugging (\ref{ordering1}) into (\ref{cor1}) we can write
\be
\label{correlation1}
\eqalign{
&\Pi(\{\tau_j \}) = \theta(\tau_2 -\tau_1) \theta(\tau_3 -\tau_4) \Pi^{(1)} (\{\tau_j \}) +
\theta(\tau_2 -\tau_1) \theta(\tau_4 -\tau_3) \Pi^{(2)} (\{\tau_j \})  \cr
&+\theta(\tau_1 -\tau_2) \theta(\tau_3 -\tau_4) \Pi^{(3)} (\{\tau_j \}) +
\theta(\tau_1 -\tau_2) \theta(\tau_4 -\tau_3) \Pi^{(4)} (\{\tau_j \}),}
\ee
where the correlation functions $\Pi^{(1,2,3,4)} (\{\tau_j \}) $ are given by
\be
\label{correlation2}
\eqalign{
\fl
 &\Pi^{(1)} (\{\tau_j \}) = \mathrm{Tr} [ \widehat{A}(\tau_1) \widehat{B} (\tau_2)  \widehat{C}(\tau_3) \widehat{D} (\tau_4)
 \hat{\rho}(\ti) ],\;\;
  \Pi^{(2)} (\{\tau_j \}) = \mathrm{Tr} [ \widehat{A}(\tau_1) \widehat{B} (\tau_2) \widehat{D} (\tau_4) \widehat{C}(\tau_3) 
 \hat{\rho}(\ti) ]  \cr
 \fl
 &\Pi^{(3)} (\{\tau_j \}) = \mathrm{Tr} [\widehat{B} (\tau_2)   \widehat{A}(\tau_1) \widehat{C}(\tau_3) \widehat{D} (\tau_4)
 \hat{\rho}(\ti) ],  \;\;
  \Pi^{(4)} (\{\tau_j \}) = \mathrm{Tr} [\widehat{B} (\tau_2) \widehat{A}(\tau_1)  \widehat{D} (\tau_4) \widehat{C}(\tau_3) 
 \hat{\rho}(\ti) ].
 }
 \ee 
Next we represent each correlation function $\Pi^{(1,2,3,4)} (\{ \tau_j \} )$ in KS functional integral formalism. 
Working out the case for $\Pi^{(1)} (\{\tau_j \})$ will be enough. 
The  operators $\widehat{A} (\tau_i), \widehat{B}(\tau_i), \widehat{C}(\tau_i), \widehat{D}(\tau_i)$ in Heisenberg picture can be expressed
in terms of the time evolution operator. For $\widehat{A}(\tau_i)$,
\be 
\label{heisenberg}
\widehat{A}(\tau_i) = \widehat{\mathsf{U}}^\dag(\tau_i,\ti) \, \widehat{A}\, 
\widehat{\mathsf{U}}(\tau_i,\ti) = \widehat{\mathsf{U}}(\ti,\tau_i) \widehat{A} \, \widehat{\mathsf{U}}(\tau_i,\ti),
\ee
and similarly for other operators.
Substituting (\ref{heisenberg}) into  $\Pi^{(1)}$ of (\ref{correlation2}) and employing the composition property of 
time evolution operator $
\widehat{\mathsf{U}}(t_1,t_3) = \widehat{\mathsf{U}}(t_1,t_2) \widehat{\mathsf{U}}(t_2,t_3) $, 
we obtain
\be
\label{correlation3}
\eqalign{
\Pi^{(1)}(\{ \tau_j \}) &=
 \mathrm{Tr} \Big [ \widehat{\mathsf{U}}(\ti, \tau_1) \, \widehat{A}\,  \U(\tau_1,\tau_2) \, 
 \widehat{B} \, \U(\tau_2, \tf) \Bigg \vert  
\U(\tf, \tau_3) \cr
& \times \widehat{C} \, \U(\tau_3,\tau_4) \widehat{D} \, \U(\tau_4,\ti) \hat{\rho}(\ti) \Big ]},
\ee
where the \textit{turning point} of the closed time contour is marked with a vertical bar.
From now on we will take a limit $\tf \to \infty, \ti \to -\infty$.
The time evolution to the left of the bar of (\ref{correlation3}) is \textit{backward} in time, 
while that  to the right of the bar is \textit{forward} in time.
Thus in (\ref{correlation3}) we have time evolution along the closed time contour, and 
 the operators are inserted along the contour.

To develop the functional integral representation,   the time evolution operators of  (\ref{correlation3}) are to be  
expressed by the Trotter product formula \cite{orland} in accordance with the
closed time contour of figure \ref{contour}.
For $ t > t'$ ( the time evolution in forward branch)
\be
\fl
\U(t,t')   =\lim_{N \to \infty } \, \underbrace{ e^{ - i   \Delta t  \, \sfH  /\hbar}  \, e^{ - i   \Delta t  \, \sfH  /\hbar} \,
\cdots e^{ - i   \Delta t  \, \sfH  /\hbar} }_{ N-{\rm times}}, \quad \Delta t = \frac{t-t'}{N}, \; \; t > t'.
\ee
And for $ t< t'$ (the time evolution in backward branch)  we note   $\U(t,t') = ( \U(t',t) )^\dag$, so that
\be
\fl
\U(t,t')   =\lim_{N \to \infty } \, \underbrace{ e^{ +i   \Delta t  \, \sfH  /\hbar} \, e^{ +i   \Delta t  \, \sfH  /\hbar} \cdots
  e^{ +i   \Delta t  \, \sfH  /\hbar}   }_{N-{\rm times}}, \quad \Delta t = \frac{t'-t}{N}, \; \; t' > t.
\ee
Then between each infinitesimal time evolution operator $e^{ \pm i   \Delta t  \, \sfH  /\hbar} $,
we insert the following  closure relation of coherent states ($\mathsf{I}$ is an  identity operator):
\be
\label{identity}
\mathsf{I} = \int D [ \{\phi_{\pm, k}\}] e^{-\phi^*_{\pm,k} \phi_{\pm,k}} \vert \{ \phi_{\pm,k} \} \ra \la \{\phi_{\pm,k} \}\vert,
\ee
where $\{ \phi_{\pm,k} \}$ and $\{\phi_{\pm,k}^* \}$ denote the collective coherent state variables
at the $k$-th time slice along the forward (+) and the backward (-) branch.
Recall that the ket coherent state is an eigenstate of annihilation operator, while the bra coherent state is an eigenstate of 
creation operator:
\be
\label{fundamental}
\hat{\mathsf{c}} \vert \phi_{\pm,k} \ra = \phi_{\pm,k} \vert \phi_{\pm,k} \ra, \quad 
\la \phi_{\pm,k} \vert \hat{\mathsf{c}}^\dagger = \la \phi_{\pm,k} \vert \phi^*_{\pm,k}.
\ee
This property makes the computation of  matrix elements of any \textit{normal-ordered} operators very easy:
\be
\label{matrix_element}
\la \phi \vert  \hat{A}_{\rm normal \; ordered} (\hat{\mathsf{c}}^\dagger, \hat{\mathsf{c}}) \vert \phi' \ra 
= e^{\phi^* \phi'}  A( \hat{\mathsf{c}}^\dagger \to \phi^*, \hat{\mathsf{c}} \to \phi').
\ee
Note that the symbol $A$ in the right hand side of (\ref{matrix_element}) is not an operator but a \textit{function} of coherent state variables
$\phi^*, \phi'$.
All operators appearing in correlation function (\ref{correlation3}) are assumed to be normal-ordered. 
Employing (\ref{identity}), (\ref{fundamental}), and (\ref{matrix_element}) it is  straightforward to develop  functional 
integral representation of the correlation function (\ref{correlation3}).
Below we exhibit a few details of the computations near operator insertion points.
Consider, for example, the insertion of the operator $\widehat{C}$ in  (\ref{correlation3}).
Clearly the time slicing can be done in such a way  that the set of time slices include the time slice at $\tau_3$ (or arbitrarily close to it).
Let  $\vert \phi_{+,k} \ra$ be the ket of the time slice corresponding to the time $\tau_3$ (recall that 
the operator $\hat{C}$ is located in the forward branch).
Then the relevant matrix elements are 
\be 
\eqalign{
& \cdots \vert\phi_{+,k+1}  \ra e^{-\phi^*_{+,k+1} \phi_{+,k+1}}
 \la\phi_{+,k+1} \vert \hat{C}  e^{-i \Delta t \sfH/\hbar}  \vert \phi_{+,k} \ra
 e^{-\phi^*_{+,k} \phi_{+,k}} \cr
 &\times \la \phi_{+,k} \vert e^{-i \Delta t \sfH/\hbar}
          \vert \phi_{+, k-1 } \ra \cdots},
\ee
Since the time interval $\Delta t$  is infinitesimally small, we can approximate 
$ e^{-i \Delta t \sfH/\hbar}   \approx 1 -i \Delta t \sfH/\hbar$ and obtain
\be
\label{intermediate}
\la\phi_{+, k+1} \vert \hat{C} \, e^{-i \Delta t \sfH/\hbar}  \vert \phi_{+,k} \ra 
 \approx \la\phi_{+,k+1} \vert \hat{C} \left[1 -i \frac{ \Delta t \sfH}{\hbar} \right ]  \vert \phi_{+,k} \ra.
\ee
Inserting a closure relation (\ref{identity}) into (\ref{intermediate}),  (\ref{intermediate}) becomes
\be
\label{inter_int}
\eqalign{
\fl
 \la\phi_{+,k+1} \vert \hat{C} &\Big[1 -i \frac{ \Delta t \sfH}{\hbar} \Big ]  \vert \phi_{+,k} \ra =
\int D[\xi^*,\xi] e^{-\xi^* \xi} \, 
\la\phi_{+,k+1} \vert \hat{C} \vert \xi \ra \la \xi \vert  \Big[1 -i \frac{\Delta t \sfH}{\hbar} \Big ]  \vert \phi_{+,k} \ra \cr
\fl
=\int D[\xi^*,\xi]& e^{-\xi^* \xi +\phi_{+,k+1}^* \xi + \xi^*\phi_{+,k}  } \, 
 C(\phi_{+,k+1}^*,\xi)  \Big[1 -i \frac{\Delta t \,\mathsf{H}(\xi^*, \phi_{+,k})}{\hbar} \Big ]  \vert \phi_{+,k} \ra.}
\ee
The operator  $\widehat{C}=\hat{C}(\hat{\mathsf{c}}^\dagger, \hat{\mathsf{c}})$
and the Hamiltonian  $\sfH$ can be assumed to be  general polynomials 
 of the creation $\hat{\mathsf{c}}^\dagger$ and  the annihilation operator $\hat{\mathsf{c}}$.
Now $\xi,\xi^*$ integrals of (\ref{inter_int})  can be done exactly.
This is because the operator $\widehat{C}$ and $\sfH$ are assumed to be polynomials of $\hat{\mathsf{c}}$ 
and $\hat{\mathsf{c}}^\dag$, so that 
the most general integrals involving $\xi$ and $\xi^*$ are of the following form: ($n,m$ are non-negative integers)
\be
\label{inte}
\int D[\xi^*,\xi] e^{-\xi^* \xi +\phi_{+,k+1}^* \xi + \xi^*\phi_{+,k}  }  \xi^n  (\xi^*)^m
=(\phi_{+,k})^n \, (\phi_{+,k+1}^*)^m.
\ee
From (\ref{inte}), it follows that (since polynomial form is preserved through integral)
\be
\label{expectation}
\fl
\la\phi_{+,k+1} \vert \hat{C}  e^{-i \Delta t \sfH/\hbar}  \vert \phi_{+,k} \ra 
\approx 
C ( \phi_{+,k+1}^*, \phi_{+,k} )\, \Big[
e^{\phi_{+,k+1}^*  \phi_{+,k}} \, e^{ -i\frac{ \Delta t }{\hbar} \mathsf{H}(\phi_{+,k+1}^*, \phi_{+,k} )}\Big ].
\ee
In the continuum limit $N \to \infty$, the difference between $\phi^*_{k+1}$ and $\phi^*_k$
in $C ( \phi_{+,k+1}^*, \phi_{+,k} )$ and $ \mathsf{H}(\phi_{+,k+1}^*, \phi_{+,k} )$ 
can be ignored.
Then (\ref{expectation}) proves that the operator $\widehat{C}(\tau)$ which is inserted in the \textit{forward} branch
maps to $C( \phi_{+}^*(\tau), \phi_{+}(\tau))$ in KS coherent functional integral formulation:
\be
\label{operator2}
\widehat{C}(\hat{\mathsf{c}}^\dagger(\tau),\hat{\mathsf{c}}(\tau))\vert_{\rm forward \; branch}
\to  C( \phi_{+}^*(\tau), \phi_{+}(\tau)).
\ee
In an entirely similar way it can be shown that
\be
\label{operator3}
\widehat{A}(\hat{\mathsf{c}}^\dagger(\tau),\hat{\mathsf{c}}(\tau))\vert_{\rm backward \; branch}  
\to  A( \phi_{-}^*(\tau), \phi_{-}(\tau)).
\ee
Combining the above results we arrive at the following KS functional integral representation of the 
correlation function (\ref{correlation3}):
\be
\label{pi1}
\eqalign{
\Pi^{(1)}(\{ \tau_j \}) &= \int D[\phi^*,\phi]\, A(\phi_-^*(\tau_1),\phi_-(\tau_1) )\,
 B(\phi_-^*(\tau_2),\phi_-(\tau_2) )  \cr
 &\times C(\phi_+^*(\tau_3),\phi_+(\tau_3) )\,
 D(\phi_+^*(\tau_4),\phi_+(\tau_4) ) \, e^{+i S_+ /\hbar  -i S_-/\hbar + \delta S_{+-}},
}
\ee
where $S_{\pm}$ is the well-known classical action of functional integral defined on the forward (+) and backward(-) branch.
$\delta S_{+-}$ is infinitesimally small terms which couple the forward and the backward branch functional integral variables.
For the subtle roles played by $\delta S_{+-}$,  see \cite{kamenev} and \cite{altland}.
We emphasize that the $A,B,C,D$ of (\ref{pi1}) are functions of ordinary variables (or anti-commuting Grassman numbers for 
fermion operators) and  that they are \textit{not} operators, so that the relative order between them does {\it not }matter.

Now recalling the definitions of the correlation functions $\Pi^{(1,2,3,4)}(\{\tau_i \})$ from  (\ref{correlation2}),
we find that 
all of $\Pi^{(1,2,3,4)}(\{\tau_i \}) $ are given by the \textit{same }expression, namely (\ref{pi1}), 
in the KS functional integral.
But they are defined 
in the separate regions of time variables $\{ \tau_i \}$ as signified by the products of step functions in
(\ref{correlation1}).
Noting a trivial identity
\be
[ \theta(\tau_1 - \tau_2) + \theta(\tau_2-\tau_1) ] \times 
[ \theta(\tau_3 - \tau_4) + \theta(\tau_4-\tau_3) ] = 1 \cdot 1 =1
\ee
we finally obtain
\be
\label{main2}
\eqalign{
\Pi(\{ \tau_i \} ) &= \mathrm{Tr}_{\rm m} \Big( \,\,\tilde{\mathtt{T}} [\widehat{A} (\tau_1) \widehat{B} (\tau_2) ] \,
                            \mathtt{T} [\widehat{C} (\tau_3) \widehat{D}(\tau_4) ] \, \hat{\rho}_{\rm m} \Big ) \cr
&= \int D[\phi^*,\phi]\, A(\phi_-^*(\tau_1),\phi_-(\tau_1) )\,
 B(\phi_-^*(\tau_2),\phi_-(\tau_2) )  \cr
 &\times C(\phi_+^*(\tau_3),\phi_+(\tau_3) )\,
 D(\phi_+^*(\tau_4),\phi_+(\tau_4) ) \, e^{+i S_+ /\hbar  -i S_-/\hbar + \delta S_{+-}}
}.
\ee
Applying the result (\ref{main2}) to the correlation function of ILS (\ref{correlation}) 
we arrive at
\be
\label{main3}
\eqalign{
\fl
&C_{k_{\rm i} \hat{\epsilon}_{\rm i}, k_{\rm f} \hat{\epsilon}_{\rm f}}(t_1,t_2,t_3,t_4)
=\int D[\phi^*,\phi] \,\, e^{+i S_+ /\hbar  -i S_-/\hbar + \delta S_{+-}}  \cr
\fl
 & \times \Big [\vec{J}_{\rm e}( \phi_-^*(\vk_{\rm i},t_1), \phi_-(\vk_{\rm i},t_1)) \cdot \hat{\epsilon}_{\rm i}^* \Big ]  
\Big [\vec{J}_{\rm e}(\phi_-^*(-\vk_{\rm f},t_2), \phi_-(-\vk_{\rm f},t_2)) \cdot \hat{\epsilon}_{\rm f} \Big ]  \cr
\fl
& \times
\Big[ \vec{J}_{\rm e}(\phi_+^*(-\vk_{\rm i},t_3),  \phi_+(-\vk_{\rm i},t_3)) \cdot \hat{\epsilon}_{\rm i} \Big ] \,  
\Big [\vec{J}_{\rm e}(\phi_+^*((\vk_{\rm f}, t_4),\phi_+((\vk_{\rm f}, t_4)) \cdot 
\hat{\epsilon}^*_{\rm f} \Big ],
}
\ee
which is the second main result of this paper.
(\ref{main3}) is the four-current correlation function which can be computed directly by using the standard 
functional integral methods such as Feynman diagram perturbation theory or semi-classical approximations.

\section{Comparison with Kramers-Heisenberg formula}
\label{comparison}

Let us compare the results (\ref{correlation}) and (\ref{cross5}) of this paper
with Kramers-Heisenberg formula which is essentially  Fermi golden rule (\ref{golden}).
At this point we emphasize that 
in deriving  (\ref{correlation}) and (\ref{cross5}) no approximations have been made regarding any
intermediate states.

Below we will point out that Kramers-Heisenberg formula was obtained by ignoring certain terms which 
are attributed  to the sudden turning on of perturbation \cite{sakurai}.
This implies that our results  (\ref{correlation}) and (\ref{cross5}) contain \textit{more} contributions than 
Kramers-Heisenberg formula, so that we have to \textit{select a subset of Feynman diagrams} contributing to 
 (\ref{correlation}) and (\ref{cross5}) which indeed
correspond to Kramers-Heisenberg formula.

We start by evaluating the time integrals of (\ref{secondM}) by introducing intermediate states $\vert \mathsf{n} \ra$.
For simplicity put $ \ti =0$,  and resolve time-ordering explicitly.
\be
\label{secondMagain}
\eqalign{
\fl
&\mathcal{M}^{(2)}_{\mathsf{FI}} =\left( -\frac{i}{\hbar} \right)^2  \int^{\tf}_{0} d \bar{t}_1  \int^{\bar{t}_1}_{0} d \bar{t}_2 
\la \mathsf{F} \vert \widehat{\mathsf{U}}_{\sfH_0} \,
\sfH_{\sfH_0}'(\bar{t}_1) \vert \mathsf{n} \ra \la \mathsf{n} \vert \sfH_{\sfH_0}'(\bar{t}_2) \vert \mathsf{I} \ra \cr
\fl
&=\frac{i}{\hbar} e^{-i \tf E_{\mathsf{F}}/\hbar} \sum_{\mathsf{n}} \,\frac{ \la \mathsf{F} \vert \sfH' \vert \mathsf{n} \ra 
\la \mathsf{n} \vert \sfH' \vert \mathsf{I} \ra}{E_{\mathsf{n}}-E_{\mathsf{I}}} \int^{\tf}_{0} d \bar{t}_1
\left( \underbrace{e^{ i \bar{t}_1 (E_{\mathsf{F}}-E_{\mathsf{I}}) /\hbar}}_{\rm Kramer- Heisenberg} - 
\underbrace{e^{ i \bar{t}_1 (E_{\mathsf{F}}-E_{\mathsf{n}}) /\hbar} }_{\rm ignored} \right )
  }
\ee
In the limit $\tf-\ti \to \infty$, only the  first term in the parenthesis of (\ref{secondMagain}) yields
a contribution which is  proportional to $\tf-\ti$ and  it implements
the energy conservation $\delta(E_{\mathsf{F}}-E_{\mathsf{I}})$.
This contribution leads to  the Fermi golden rule (\ref{golden}) and to the Kramers-Heisenberg formula.
In the same limit  the second term does not give substantial contribution unless $E_{\mathsf{F}} \sim E_{\mathsf{n}}$, and 
this constraint
greatly suppresses the number of possible intermediate states $\vert \mathsf{n} \ra$, resulting in negligible contribution compared 
to the first one \cite{sakurai}.
The constraint $E_{\mathsf{F}} \sim E_{\mathsf{n}}$ can be enforced by introducing 
$\delta(E_{\mathsf{F}}- E_{\mathsf{n}})$ in the sum over intermediate states.
This delta function can be naturally understood as a \textit{spectral function}  which is the imaginary part of retarded Green's function.
Therefore, in our formalism, we have to ignore the Feynman diagrams which have 
\textit{more spectral functions than those which correspond to  Kramers-Heisenberg formula}.
 This assertion will be explicitly demonstrated in section \ref{application}.
In transition {\it probability},  the  \textit{ignored} terms (which is to be  squared) 
will allow processes which are not constrained by the energy conservation
$E_{\mathsf{F}} = E_{\mathsf{I}}$. These processes will turn out to be  strongly suppressed by more factors of spectral function.
An explicit example will be given in section \ref{application}.

So this is the price we have to pay: we have obtained a correlation function which does not depend on intermediate states,
but  we have to select appropriate Feynman diagrams.
Fortunately, the selection of diagrams will turn out to be  more or less obvious and straightforward.

We also note that the \textit{ignored} terms can be  eliminated 
by adiabatic turning on of  perturbations \cite{sakurai}, but in our formalism this method turns out to be very cumbersome, 
so it is not pursued further.

\section{Application: single G-phonon Raman intensity of graphene}
\label{application}
In this section we apply  KS functional integral formalism  to the case of the  single G-phonon Raman intensity of graphene 
and compare the results  with those obtained  by
conventional approach \cite{basko} based on Fermi golden rule.

For explicit calculations we need more elaborations in KS formalism.
More details  can be found in references \cite{kamenev,altland}.
For fermions let us use the notation $(\psi_\pm, \psi^*_\pm)$ for coherent state variables instead of $(\phi_\pm,\phi^*_\pm)$, and  
reserve the notation $(\phi_\pm,\phi^*_\pm)$ for bosons.
It turns out that  in practical calculations it is more convenient to take a linear combination of coherent state variables  
(often called Keldysh basis)
 in the following way \cite{kamenev,altland}. 
  For \textit{bosons}, the Keldysh basis is given by ( ``cl''=classical,  ``q''= quantum,  see \cite{kamenev} )
 \be
 \eqalign{
 \fl
 \phi_{\rm cl}(\vk,t) = \frac{1}{\sqrt{2}} \Big [ \phi_+(\vk,t) +  \phi_-(\vk,t)  \Big ], \quad
 \phi_{\rm q}(\vk,t) = \frac{1}{\sqrt{2}} \Big [ \phi_+(\vk,t) -  \phi_-(\vk,t)  \Big ], \cr
 \fl
  \phi_{\rm cl}^*(\vk,t) = \frac{1}{\sqrt{2}} \Big [ \phi_+^*(\vk,t) +  \phi_-^*(\vk,t)  \Big ], \quad
 \phi_{\rm q}^*(\vk,t) = \frac{1}{\sqrt{2}} \Big [ \phi_+^*(\vk,t) -  \phi_-^*(\vk,t)  \Big ]. }
 \ee
 For \textit{ fermions}, a different combination is chosen.
 \be
 \label{fermion}
 \eqalign{
 \fl
 \psi_1(\vk,t) &= \frac{1}{\sqrt{2}} \Big[\psi_+(\vk,t) + \psi_-(\vk,t) \Big ], \quad 
 \psi_2(\vk,t) = \frac{1}{\sqrt{2}} \Big[ \psi_+(\vk,t) - \psi_-(\vk,t) \Big ], \cr
 \fl
 \bar{\psi}_1(\vk,t) &= \frac{1}{\sqrt{2}} \Big [\psi_+^*(\vk,t) - \psi_-^*(\vk,t) \Big ], \quad 
 \bar{\psi}_2(\vk,t) = \frac{1}{\sqrt{2}} \Big [\psi_+^*(\vk,t) +\psi_-^*(\vk,t) \Big ]. }
 \ee
Then Green's functions of bosons (such as phonon) are given by ($a,b = {\rm cl, q}$ )
\be
-i \la \phi_a(\vk,t) \phi^*_b(\vk,t') \ra = \widehat{D}^{ab}_{\vk}(t,t') =  
\left( \matrix{ D^{\rm K}_{\vk}(t,t')  &  D^{\rm R}_{\vk}(t,t') \cr
            D^{\rm A}_{\vk}(t,t')  & 0  } \right ),
\ee
where the superscripts ``K,R,A'' indicate Keldysh, retarded, and advanced Green's functions, respectively.
On the other hand.  Green's functions of fermions are given by ($a,b=1,2$)
\be
-i \la \psi_a(\vk,t) \bar{\psi}_b(\vk,t') \ra = \widehat{G}^{ab}_{\vk}(t,t') =  
\left( \matrix{ G^{\rm R}_{\vk}(t,t')  &  G^{\rm K}_{\vk}(t,t') \cr
              0 &  G^{\rm A}_{\vk}(t,t') } \right ).
\ee
At equilibrium, the Keldysh Green's function is determined by retarded and advanced 
Green's functions owing to the fluctuation-dissipation theorem \cite{kamenev,rammer}.
\be
\label{keldysh}
\eqalign{
D^{\rm K}(\omega) &= \coth \frac{ \omega}{2 k_B T}\,\, \Big[ D^{\rm R}(\omega) - D^{\rm A}(\omega) \Big]: \;{\rm bosons},\cr
G^{\rm K}(\epsilon) &= \tanh \frac{ \epsilon}{2 k_B T}\,\, \Big[ G^{\rm R}(\epsilon) - G^{\rm A}(\epsilon) \Big]: \;{\rm fermions}.}
\ee
Since the retarded and the advanced Green's functions in frequency space are complex conjugate of each other, 
(\ref{keldysh}) implies 
that the Keldysh Green's function is \textit{purely imaginary} (or anti-Hermitian).
We recall the difference between the retarded and the advanced Green's function in energy space is $2i$ times the spectral function,
which means that in equilibrium the Keldysh Green's function is  always proportional to the spectral function.
\begin{figure}[!hbt]
\begin{center}
\includegraphics[width = 0.7 \textwidth]{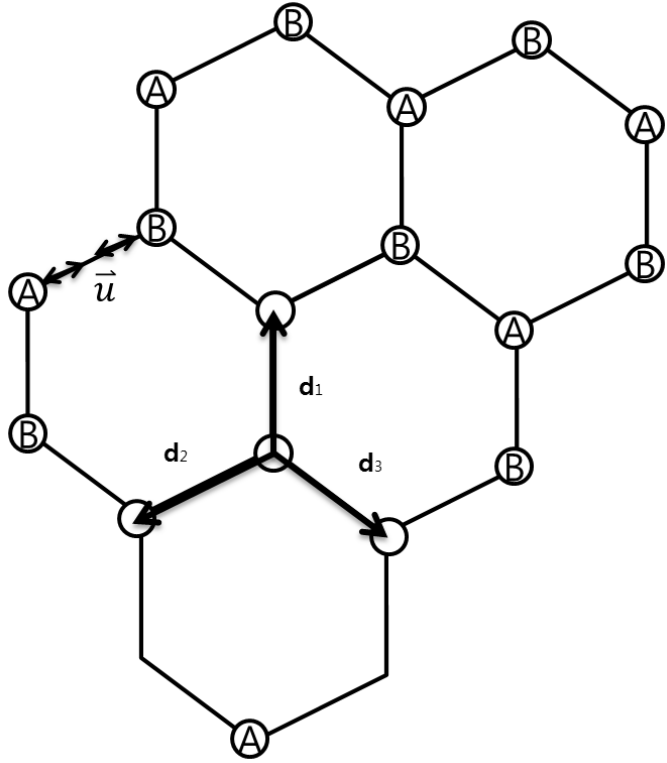}
\caption{ The honeycomb lattice structure of graphene with two sublattices A and B. 
$\mathbf{d}_{i=1,2,3}$ are bond vectors connecting three nearest-neighbor carbon atoms. 
The phonon mode $\vec{u}$ is also depicted. }
\label{lattice}
\end{center}
\end{figure}

Now we turn to the case of graphene [for a review, see \cite{neto}]. 
The graphene has honeycomb lattice structure with two sublattices [denoted by ${\rm A,B}$ in figure \ref{lattice} ], so that
all the electron Green's functions become 2x2 matrix in sublattice space (the spin will be neglected for simplicity).
More concretely,
\be
G^{\rm R}_{\vk}(\epsilon)= \Big[ ( \epsilon + i 0^+ ) \mathsf{I}_2 - \big[ \matrix{
 0 & t_\vk \cr  t_\vk^*  & 0 } \big ] \Big]^{-1}, \;  G^{\rm A}_{\vk}(\epsilon) =[ G^{\rm R}_{\vk}(\epsilon)  ]^{\dagger},
 \ee
where $t_\vk =- t_0 \sum_{i=1,2,3} e^{- i \vk \cdot \mathbf{d}_i}$ with $\mathbf{d}_i$ being vectors connecting nearest neighbors and
$t_0$ is a hopping amplitude. $\mathsf{I}_2$ is the identity matrix in sublattice space.
The retarded Green's function of  G-phonon ( $\vec{u}$ in figure \ref{lattice})  is given by  ($M$ is the mass of  carbon atom)
\be
\label{gphonon}
D^{\rm R}_{\vq}(\omega)=  \frac{1}{M} \frac{1}{(\omega+ i 0^+)^2 - \omega_\vq^2}, \;\;
D^{\rm A}_{\vq}(\omega)=[ D^{\rm R}_{\vq}(\omega)]^*,
\ee
where $\omega_\vq$ is the G-phonon dispersion.
The Einstein phonon approximation $\omega_\vq \approx  \omega_{\rm ph}$ will be assumed.

The  electric  currents of (\ref{main3})  of graphene \cite{basko} can be found
in terms of coherent state variables using (\ref{fermion}):
\be
\label{current}
\vec{J}_{\rm \pm}(\vq) =  (-e)\frac{1}{2} \sum_\vk  \bar{\Psi}(\vk-\vq) \, \mathsf{E}_\pm
\otimes \hat{V}_{\vk-\vq/2} \Psi(\vk),
\ee
where $\vec{v}_\vk = \frac{\partial t_\vk}{\partial \vk}$, 
$\Psi= ( \psi_{1 A}, \psi_{1 B}, \psi_{2 A}, \psi_{2 B} )^{\rm t}$, and
\be
\mathsf{E}_\pm =  \left( \matrix{ \pm 1&  1 \cr  1 & \pm 1 } \right ), \quad
 \hat{V}_{\vk-\vq/2}  =  \left( \matrix{ 0 & \vec{v}_{\vk-\vq/2} \cr \vec{v}^*_{\vk-\vq/2} & 0 } \right ).
\ee
For the Raman scattering of graphene, the photon momentum can be neglected, so that $\vq \approx 0$.
Finally, the (simplified) interaction between electron and G-phonon  in action form is given by ($N$ is the number of lattice sites)
\be
\fl
S_{\rm e-ph}  = -\lambda \frac{1}{\sqrt{N}} \int dt  \sum_{\vq,\vk} \bar{\Psi}(\vk+\vq) 
( u^{\rm cl}_\vq \mathrm{I}_2 + u ^{\rm q}_\vq  \sigma_x ) \otimes \hat{f}_{\vk,\vq}
\Psi(\vk),
\ee
where $ \gamma^{\rm cl} =\mathrm{I}_2$ (identity matrix) and $\gamma^{\rm q} = \sigma_x$ (Pauli matrix) act on the Keldysh space
and $\hat{f}_{\vk,\vq} =   \left ( \matrix{ 0 & f_{\vk,\vq} \cr f^*_{\vk,\vq} & 0 } \right )$.
$u^{\rm cl,q}=\frac{1}{\sqrt{2 M \omega_\vq}} [ \phi^{\rm cl,q}(\vq) + \phi^{\rm * cl,q}(-\vq) ]$
is the phonon coordinate in Keldysh basis and $\lambda$ is the electron-phonon coupling constant.
The detailed forms of  functions $v_{\vk}$ and $f_{\vk,\vq}$ do not concern us in this paper.

\begin{figure}[!hbt]
\begin{center}
\includegraphics[width = 0.9 \textwidth]{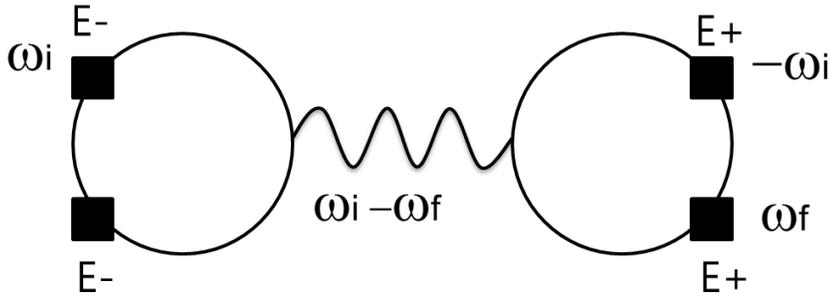}
\caption{The lowest order Feynman diagram for one-phonon Raman scattering.
The solid lines and the way line  represent the  electron  KS Green's function and the phonon KS Green's function,
respectively.
The black boxes indicate the insertion of current operators of the correlation function (\ref{main3}).}
\label{feyn1}
\end{center}
\end{figure}

Now we are ready to compute the single G-phonon Raman intensity graphene in KS functional integral  formalism.
The single phonon Raman intensity in the lowest order perturbation theory  is described by the  Feynman diagram of 
figure \ref{feyn1}.
The direct application of  Feynman rules yields the following result for the correlation function (\ref{main3}):
\be
\label{final}
\tilde{C}_{\omega_{\rm i} \hat{\epsilon}_{\rm i},  \omega_{\rm f} \hat{\epsilon}_{\rm f}}
= \lambda^2 \, M^a_{--}  \,  \big [ i D^{ab}_{\vq =0} (\omega_{\rm i} - \omega_{\rm f}) \big ]\,M^b_{++}, \;\;
a,b={\rm cl,q},
\ee
where ($M^a_{--}$ is for the  loop in the left side of Fig. \ref{feyn1}, and $M^b_{++}$ is for the loop in the right)

\be
\label{dia1}
\eqalign{
\fl
 M^a_{--} &= \sum_\vk \int^\infty_{-\infty} \frac{d \epsilon}{2\pi} 
 \Bigg \{ \mathrm{Tr} \left [ \widehat{G}_\vk(\epsilon) ( \mathsf{E}_- \otimes \hat{V}_\vk ) 
 \widehat{G}_\vk(\epsilon+\omega_{\rm i} ) (\gamma^a \otimes \hat{f}_{\vk,\vq=0}) \widehat{G}_\vk(\epsilon+\omega_{\rm f}) 
  ( \mathsf{E}_- \otimes \hat{V}_\vk ) \right ]  \cr
  \fl
  &+
\mathrm{Tr} \left [ \widehat{G}_\vk(\epsilon) ( \mathsf{E}_- \otimes \hat{V}_\vk ) 
 \widehat{G}_\vk(\epsilon-\omega_{\rm f} ) (\gamma^a \otimes \hat{f}_{\vk,\vq=0}) \widehat{G}_\vk(\epsilon-\omega_{\rm i}) 
  ( \mathsf{E}_- \otimes \hat{V}_\vk ) \right ]    \Bigg \}
  }\ee
  \be
  \label{dia2}
  \eqalign{
 \fl
 M^b_{++} &= \sum_\vk \int^\infty_{-\infty} \frac{d \epsilon}{2\pi} 
 \Bigg \{ \mathrm{Tr} \left [ \widehat{G}_\vk(\epsilon) ( \mathsf{E}_+ \otimes \hat{V}_\vk ) 
 \widehat{G}_\vk(\epsilon-\omega_{\rm i} ) (\gamma^b \otimes \hat{f}_{\vk,\vq=0}) \widehat{G}_\vk(\epsilon-\omega_{\rm f}) 
  ( \mathsf{E}_+ \otimes \hat{V}_\vk ) \right ]  \cr
  \fl
  &+
\mathrm{Tr} \left [ \widehat{G}_\vk(\epsilon) ( \mathsf{E}_+ \otimes \hat{V}_\vk ) 
 \widehat{G}_\vk(\epsilon+\omega_{\rm f} ) (\gamma^b \otimes \hat{f}_{\vk,\vq=0}) \widehat{G}_\vk(\epsilon+\omega_{\rm i}) 
  ( \mathsf{E}_+\otimes \hat{V}_\vk ) \right ]    \Bigg \}.  
  }
 \ee
The trace ``Tr'' is over both  Keldysh and  sublattice spaces.
The evaluation of the correlation function proceeds as follows:
(1) take the trace over Keldysh space
(2) then  using the  (anti) Hermiticity properties  of matrices of Green's functions  it is easily shown that
 $ M^{\rm cl}_{--}= (M^{\rm cl}_{++})^*$ and $M^{\rm q}_{--}= - (M^{\rm q}_{++})^*$
(3) sort out all terms with more than two Keldysh Green's function $G^{\rm K}$ or spectral functions;
these terms belong to  the \textit{ignored} terms discussed in section \ref{comparison}. Direct integration indeed shows that
there is no resonant behaviour (roughly \textit{two} delta functions eliminate all singular behaviours).
(4) \textit{after} the sorting, a relation $ M^{\rm cl}_{--} = -M^{\rm q}_{--}$ can be shown  by  using the
Hermiticity of matrices and the property
$ \int d \epsilon G^{\rm R}(\epsilon) G^{\rm R}(\epsilon \pm \omega_{\rm i/f}) G^{\rm R}(\epsilon \pm \omega_{\rm f/i})=0$
(also for the advanced Green's function)
which follows from the analyticity of the retarded Green's functions
(5) finally using the property $ [G^{\rm R}]^\dag = G^{\rm A}$ and $ [G^{\rm K}]^\dag =- G^{\rm K}$, $M^{\rm cl}_{--}$ can be shown 
to be purely \textit{imaginary}.
Then the correlation function boils down to 
\be
\label{final1}
\eqalign{
\fl
& \tilde{C}_{\omega_{\rm i} \hat{\epsilon}_{\rm i},  \omega_{\rm f} \hat{\epsilon}_{\rm f}}=\lambda^2
[- (M^{\rm cl}_{--})^2 ] i \Big [  D^{\rm K}(\omega_{\rm i} - \omega_{\rm f}) +D^{\rm R}(\omega_{\rm i} 
- \omega_{\rm f})-D^{\rm A}(\omega_{\rm i} - \omega_{\rm f}) \Big ] \cr
\fl
&=\lambda^2
[- (M^{\rm cl}_{--})^2 ] \Big [  \coth \left( \frac{ \omega_{\rm i} -\omega_{\rm f}}{2 k_B T} \right ) +1 \Big ]
i \Big[ D^{\rm R}(\omega_{\rm i} 
- \omega_{\rm f})-D^{\rm A}(\omega_{\rm i} - \omega_{\rm f}) \Big ] \cr
\fl
&=\frac{\lambda^2}{\pi M \omega_{\rm ph}}
[- (M^{\rm cl}_{--})^2 ] \Big [  \coth \left( \frac{ \omega_{\rm i} -\omega_{\rm f}}{2 k_B T} \right ) +1 \Big ]
\Big[ \delta(\omega_{\rm i} - \omega_{\rm f} - \omega_{\rm ph}) -
\delta(\omega_{\rm i} - \omega_{\rm f} + \omega_{\rm ph}) \Big ], }
\ee
where (\ref{keldysh}) is employed in the second line and the explicit form of phonon Greens' function (\ref{gphonon}) is used 
in the third line.
The delta function $\delta(\omega_{\rm i} - \omega_{\rm f} - \omega_{\rm ph}) $ of (\ref{final1}) corresponds to the 
Stokes process of phonon emission which dominates over the anti-Stokes process of phonon absorption 
in the low temperature limit due to the thermal occupation number factor 
$  \coth [(\omega_{\rm i} -\omega_{\rm f})/2 k_B T ] +1$.
Now the computation of $M^{\rm cl}_{--}$ is rather straightforward (but tedious).
Substituting (\ref{final1}) into (\ref{cross5}), the integral over the final photon momentum $\vk_{\rm f}$ is done by the 
delta function of (\ref{final1}), leaving behind an integral over the solid angle of $\vk_{\rm f}$, which defines a 
\textit{differential} cross section.
The result obtained in this way  is identical with the equations (7), (12a) of \cite{basko} [$F_0 \alpha_3$ part of (12b,c,d)]. 
More precisely, \cite{basko} also includes additional phonon-photon coupling specific to graphene  which is irrelevant
in our discussion.

\begin{figure}[!hbt]
\begin{center}
\includegraphics[width = 0.7 \textwidth]{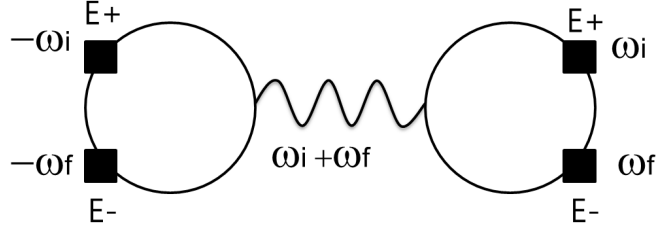}
\caption{The Feynman diagram which does not satisfy \textit{correct} energy conservation but is allowed by Feynman rule.}
\label{feyn2}
\end{center}
\end{figure}

The Feynman diagram of figure \ref{feyn2} depicts a process for the correlation function (\ref{main3})  which is allowed by Feynman rule 
but does not satisfy the \textit{correct} energy conservation, as can be seen clearly  in the  phonon energy  
$\omega_{\rm i} + \omega_{\rm f}$.
This Feynman diagram corresponds to the squared ignored terms discussed in section \ref{comparison}.
Direct computation shows that all terms contributing to this Feynman diagrams contain at least two factors of 
spectral functions or Keldysh Green function, so they are strongly suppressed and do not contribute to the resonant scattering.
A simple criterion for this type of diagrams  is that they have 
$\mathsf{E}_+$ and $\mathsf{E}_-$ vertices in the same loop (compare with figure \ref{feyn1}). 

\section{Inelastic light scattering for systems in nonequilibrium conditions}
\label{noneq}
 KS functional integral approach for ILS  can be readily generalized to the systems in  non-equilibrium conditions.
In fact, the study of non-equilibrium phenomena was the original motivation for the introduction of  KS method. 
We will assume that the  non-equilibrium conditions are created by certain driving forces for the matter systems.
Specifically 
we will consider the semiconductors of direct transition type such as GaAs, where the driving forces are 
\textit{pump pulse laser field} which can be treated classically.
This pump pulse field \textit{prepares the initial matter state} $ \vert \mathrm{i} \ra$ at time $\ti$ for ILS,
and then the probe photon of ILS is incident \cite{shah}.

In the presence of the pumping field,  the time translation invariance is certainly violated, so that the 
definition of differential cross-section (\ref{cross}) is modified in the following way:
\be
\label{cross-noneq}
d \sigma = \frac{d}{d t_{\mathsf{F}}} \,\sum_{\mathsf{I}} p_{\mathsf{I}}    \sum_{\mathsf{F}} 
 \frac{| \mathcal{ M}_{\mathsf{FI}}(t_{\mathsf{F}},t_{\mathsf{I}}) |^2}{c/\mathcal{V}}.
\ee
The initial matter state $\vert \mathrm{i} \ra$ is prepared from the remote past  (say $-\infty$) state [which is  denoted as
$ \vert \mathrm{i}_{-\infty} \ra$ ] 
by the action of the pumping field and the interactions in matter systems.
The pumping field is assumed to be turned off before the probe photon  comes in,
$  \widehat{\mathsf{H}}_{\rm pump} (t) = 0 $ for $t > t_{\mathsf{I}}$.
Thus we have
\be
\vert \mathrm{i} \ra = \widehat{\mathsf{U}}_{\widehat{\mathsf{H}}_{\rm m} + \widehat{\mathsf{H}}_{\rm pump}}(t_{\mathsf{I}},-\infty)\,
\vert \mathrm{i}_{-\infty} \ra,
\ee
where 
\be
\label{evolution-1}
 \widehat{\mathsf{U}}_{\widehat{\mathsf{H}}_{\rm m}+\widehat{\mathsf{H}}_{\rm pump}
 }(t_{\mathsf{I}},-\infty)  =\mathtt{T} 
 \exp\Big[-\frac{i}{\hbar} \int^{t_{\mathsf{I}}}_{-\infty} \, 
 d \bar{t} \, 
 \big (\widehat{\mathsf{H}}_{\rm m}(\bar{t})+\widehat{\mathsf{H}}_{\rm pump}(\bar{t})\big )\Big ].
\ee
The distribution of the remote past states are descrbied by density matrix:
\be
p_{\mathrm{i}_{-\infty}} = \la \mathrm{i}_{-\infty} \vert \hat{\rho}_{\rm m,-\infty} \vert \mathrm{i}_{-\infty} \ra.
\ee
Now the equation (\ref{cross2-1}) is modified to:
\be
\label{cross2-2}
\eqalign{
\fl
&\sum_{\rm f} \vert \mathcal{M}^{(2)}_{\mathsf{FI}} \vert^2 = 
 \frac{1}{(2!)^2 \hbar^4} \,\prod_{l=1}^4 \int_{\ti}^{\tf} d \bar{t}_l  \cr
 &\fl
\la {\rm i}_{-\infty} \vert \widehat{\mathsf{U}}^\dag_{
\widehat{\mathsf{H}}_{\rm m} + \widehat{\mathsf{H}}_{\rm pump}}(t_{\mathsf{I}},-\infty)\widehat{\mathsf{D}}^\dag_{\tilde{k}_{\rm i},\tilde{k}_{\rm f}}(\bar{t}_3,\bar{t}_4)  
\widehat{\mathsf{D}}_{\tilde{k}_{\rm i},\tilde{k}_{\rm f}}(\bar{t}_1,\bar{t}_2)
\widehat{\mathsf{U}}_{
\widehat{\mathsf{H}}_{\rm m} + \widehat{\mathsf{H}}_{\rm pump}}(t_{\mathsf{I}},-\infty)\vert {\rm  i}_{-\infty} \ra.
}
\ee
Then the differential cross-section takes the form
\be
\label{cross3-1}
\eqalign{
\fl
& d \sigma = \frac{1}{(2!)^2 \hbar^4}\frac{1}{c/\mathcal{V}}\frac{d }{d\tf} \,  \sum_{\vk_{\rm f},\hat{\epsilon}_{\rm f}} \,
\prod_{l=1}^4 \int_{\ti}^{\tf} d \bar{t}_l \cr
& \fl \mathrm{Tr}_{\rm m} \Big[ \hat{\rho}_{\rm m,-\infty} \widehat{\mathsf{U}}_{
\widehat{\mathsf{H}}_{\rm m} + \widehat{\mathsf{H}}_{\rm pump}}(-\infty,t_{\mathsf{I}})\,
  \widehat{\mathsf{D}}^\dag_{\tilde{k}_{\rm i},\tilde{k}_{\rm f}}(\bar{t}_3,\bar{t}_4) 
  \widehat{\mathsf{D}}_{\tilde{k}_{\rm i},\tilde{k}_{\rm f}}(\bar{t}_1,\bar{t}_2)
  \widehat{\mathsf{U}}_{
\widehat{\mathsf{H}}_{\rm m} + \widehat{\mathsf{H}}_{\rm pump}}(t_{\mathsf{I}},-\infty)\Big ].}
\ee
The expressions for the operators $ \widehat{\mathsf{D}}_{\tilde{k}_{\rm i},\tilde{k}_{\rm f}}$
and $ \widehat{\mathsf{D}}^\dag_{\tilde{k}_{\rm i},\tilde{k}_{\rm f}}$[ the equations (\ref{op1},\ref{op2}) ]
remain unchanged since the pumping field is turned off after $t_{\mathsf{I}}$.
Substituting (\ref{op1},\ref{op2}) into (\ref{cross3-1}) and defining a correlation function 
\be
\label{correlation-1}
\fl
\eqalign{
\tilde{C}_{k_{\rm i} \hat{\epsilon}_{\rm i}, k_{\rm f} \hat{\epsilon}_{\rm f}}(t_3,t_4,t_1,t_2)
&=\mathrm{Tr}_{\rm m} \Big[ \hat{\rho}_{\rm m,-\infty} 
\widehat{\mathsf{U}}_{
\widehat{\mathsf{H}}_{\rm m} + \widehat{\mathsf{H}}_{\rm pump}}(-\infty,t_{\mathsf{I}})\,
\tilde{\mathtt{T}} [ \vec{J}_{\rm e}(\vk_{\rm i}, t_3) \cdot \hat{\epsilon}_{\rm i}^*  \, \,
\vec{J}_{\rm e}(-\vk_{\rm f}, t_4) \cdot \hat{\epsilon}_{\rm f}  ] \cr
&\times \mathtt{T} [ \vec{J}_{\rm e}(-\vk_{\rm i}, t_1) \cdot \hat{\epsilon}_{\rm i} \,  \,\vec{J}_{\rm e}(\vk_{\rm f}, t_2) \cdot 
\hat{\epsilon}^*_{\rm f} ] 
\widehat{\mathsf{U}}_{
\widehat{\mathsf{H}}_{\rm m} + \widehat{\mathsf{H}}_{\rm pump}}(\ti,-\infty)\,\Big ].
}
\ee
we arrive at
\be
\label{cross4-1}
\fl
d \sigma =\frac{(2\pi/\hbar)^2}{c  \omega_{\rm i} \omega_{\rm f}} \, \frac{d}{d\tf} 
\, \sum_{\vk_{\rm f},\hat{\epsilon}_{\rm f}} \int^{\tf}_{\ti} (d t_i) \,
e^{-i \omega_{\rm i} t_1 + i \omega_{\rm f} t_2 + i \omega_{\rm i} t_3 - i\omega_{\rm f} t_4}
\tilde{C}_{k_{\rm i} \hat{\epsilon}_{\rm i}, k_{\rm f} \hat{\epsilon}_{\rm f}}(t_3,t_4,t_1,t_2),
\ee

The equation (\ref{cross4-1}) is the differential cross section of ILS for  systems in non-equilibrium conditions
in the form of correlation function.

Next we disentangle the time ordering of the correlation function
(\ref{correlation-1}) by employing KS functional integral formulation as in
Section \ref{functional}.
The relevant correlation function is of the form ( in the same notations as those of the section \ref{functional} )
\be
\fl
\tilde{\Pi}( \{ \tau_i \} )=
\mathrm{Tr}_{\rm m} \Big[  \widehat{\mathsf{U}}_{
\widehat{\mathsf{H}}_{\rm m} + \widehat{\mathsf{H}}_{\rm pump}}(-\infty,t_{\mathsf{I}})\,
\tilde{\mathtt{T}} [\widehat{A} (\tau_1) \widehat{B} (\tau_2) ] \,
                            \mathtt{T} [\widehat{C} (\tau_3) \widehat{D}(\tau_4) ] \,  
  \widehat{\mathsf{U}}_{
\widehat{\mathsf{H}}_{\rm m} + \widehat{\mathsf{H}}_{\rm pump}}(t_{\mathsf{I}},-\infty)\hat{\rho}_{\rm m}\Big ].
\ee
The difference with the equilibrium case (\ref{cor1}) lies in the presence of the 
pumping factor 
$\widehat{\mathsf{U}}_{\widehat{\mathsf{H}}_{\rm m} + \widehat{\mathsf{H}}_{\rm pump}}$.
Next we follow the same steps as (\ref{ordering1}, \ref{correlation1}, \ref{correlation2}).
Employing the composition property of the time evolution operator we obtain
\be
\fl
\label{correlation3-1}
\eqalign{
\tilde{\Pi}^{(1)}(\{ \tau_j \}) &=
 \mathrm{Tr} \Big [\widehat{\mathsf{U}}_{\widehat{\mathsf{H}}_{\rm m} + \widehat{\mathsf{H}}_{\rm pump}}(
 -\infty,\ti)\,
 \widehat{\mathsf{U}}_{\widehat{\mathsf{H}}_{\rm m}}(\ti, \tau_1) \, \widehat{A}\,  
 \U_{\widehat{\mathsf{H}}_{\rm m}}(\tau_1,\tau_2) \, 
 \widehat{B} \, \U_{\widehat{\mathsf{H}}_{\rm m}}(\tau_2, \tf) \Bigg \vert  
\U_{\widehat{\mathsf{H}}_{\rm m}}(\tf, \tau_3) \cr
& \times \widehat{C} \, \U_{\widehat{\mathsf{H}}_{\rm m}}(\tau_3,\tau_4) \widehat{D} \, 
\U_{\widehat{\mathsf{H}}_{\rm m}}(\tau_4,\ti) 
\widehat{\mathsf{U}}_{\widehat{\mathsf{H}}_{\rm m} + \widehat{\mathsf{H}}_{\rm pump}}(\ti,-\infty)
\hat{\rho}(-\infty) \Big ]},
\ee
Since the pumping field is turned off   in the time interval $[\ti,\tf]$, the subscripts 
denoting the Hamiltonian governing time evolution are, in fact,  redundant. Namely we can simply
take the total Hamiltonian to be $\widehat{\mathsf{H}}_{\rm m} + \widehat{\mathsf{H}}_{\rm pump}$
with the understanding that the pumping field vanishes for $t > \ti$.
Then again by the composition property of evolution operator, 
the equation (\ref{correlation3-1}) reduces to 
\be
\fl
\label{correlation3-2}
\eqalign{
\tilde{\Pi}^{(1)}(\{ \tau_j \}) &=
 \mathrm{Tr} \Big [\widehat{\mathsf{U}}(
 -\infty,\tau_1)\, \widehat{A}\,  
 \U(\tau_1,\tau_2) \, 
 \widehat{B} \, \U(\tau_2, \infty) \Bigg \vert  
\U(+\infty, \tau_3) \cr
& \times \widehat{C} \, \U(\tau_3,\tau_4) \widehat{D} \, 
\U(\tau_4,-\infty)
\hat{\rho}(-\infty) \Big ]},
\ee
where we have inserted an identity $
\U(\tf,+\infty) \U(+\infty,\tf) = \mathrm{I}$.
Now the equation (\ref{correlation3-2}) is of the same form as that of (\ref{correlation3}), so that
they have the identical form of the following functional integral representation:
\be
\label{main4}
\eqalign{
\tilde{\Pi}(\{\tau_i \} ) &  = \int D[\phi^*,\phi]\, A(\phi_-^*(\tau_1),\phi_-(\tau_1) )\,
 B(\phi_-^*(\tau_2),\phi_-(\tau_2) )  \cr
 &\times C(\phi_+^*(\tau_3),\phi_+(\tau_3) )\,
 D(\phi_+^*(\tau_4),\phi_+(\tau_4) ) \, e^{+i S_+ /\hbar  -i S_-/\hbar + \delta S_{+-}}
},
\ee
where the actions $S_\pm$ now include the driving pumping field term.
\be
\label{action-noneq}
S_\pm [ \phi_\pm ] =  \int^\infty_{-\infty} d t \Big[ i \phi^*_\pm \hbar \partial_t  \phi_\pm
-H_{\rm m} (\phi_\pm)- H_{\rm pump}(\phi_\pm) \Big ].
\ee
With the equations (\ref{main4},\ref{action-noneq}),  ILS in non-equilibrium situation can be
computed by using the standard functional integral method.

Next let us apply the above results to the case of ILS of the semiconductor pumped by intense  laser pulse.
The pumping Hamiltonian in the dipole approximation
of optical interband transition \cite{Haug-Koch} is given by 
[ for simplicity  two band approximation  (one conduction
and one valence band ) is made and spin degrees of freedom is ignored, see Chapter 5 of \cite{Haug-Koch} ]
\be
\label{dipole}
\sfH_{\rm pump} = - \mathcal{E}(t) \sum_\vk ( d_{cv} \hat{\mathsf{c}}^\dag_{c,\vk} \hat{\mathsf{c}}_{v,\vk}+ 
d_{cv}^* \hat{\mathsf{c}}^\dag_{v,\vk} \hat{\mathsf{c}}_{c,\vk} ),
\ee
where $\mathcal{E}(t)$ is the electric field of the pumping pulse laser
and $d_{cv}$ is the dipole matrix element.
$\hat{\mathsf{c}}_{c,v}$ is the electron annihilation operator for conduction and valence band, respectively.

The electric current operator in dipole approximation is given by 
 [$w_{\vk,\vq}$ is the
optical transition matrix element]
\be
\label{xxx}
\vec{J}_{\rm e}(\vq) = \sum_\vk \Big ( w_{\vk, \vq} \hat{\mathsf{c}}^\dag_{c \vk+\vq} \hat{\mathsf{c}}_{v \vk } + 
 w^*_{\vk,\vq} \hat{\mathsf{c}}^\dag_{v \vk+\vq} \hat{\mathsf{c}}_{c \vk} \Big ).
 \ee
In terms of the coherent state variables in Keldysh basis  (\ref{xxx}) becomes
\be
\label{current-1}
\vec{J}_{\rm e, \pm}(\vq) = \frac{1}{2} \sum_\vk  \bar{\Psi}(\vk-\vq) \, \mathsf{E}_\pm
\otimes \hat{V}_{\vk-\vq/2} \Psi(\vk),
\ee
where  
$\Psi= ( \psi_{1 c}, \psi_{1 v}, \psi_{2 c}, \psi_{2 v} )^{\rm t}$, and
\be
\mathsf{E}_\pm =  \left( \matrix{ \pm 1&  1 \cr  1 & \pm 1 } \right ), \quad
 \hat{V}_{\vk-\vq/2}  =  \left( \matrix{ 0 & \vec{w}_{\vk-\vq/2} \cr \vec{w}^*_{\vk-\vq/2} & 0 } \right ).
\ee
Then the action for the pumping term is given by
\be
\label{pumpaction}
S_{\rm pump} = \int^\infty_{-\infty} dt \, \mathcal{E}(t) \sum_\vk  \bar{\Psi}_\vk(t) \mathrm{I} \otimes 
\hat{d} \Psi_\vk(t),
\ee
where
$\hat{d}=  \left(
 \matrix{
  0 & d_{cv} \cr d^*_{cv} & 0 
 } \right )$.
 
\begin{figure}[!hbt]
\begin{center}
\includegraphics[width = 0.7 \textwidth]{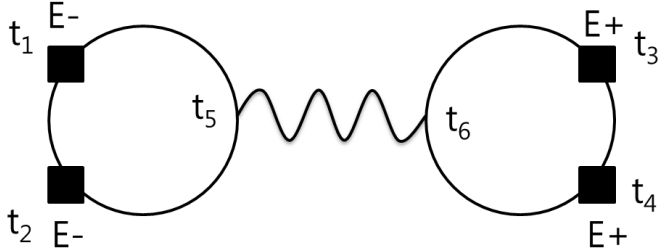}
\caption{The lowest Feynman diagram for one-phonon Raman scattering in non-equilibrium condition. Note that the solid lines now 
represent the electron Green function \textit{in non-equilibrium} (\ref{zz2}). }
\label{realtime}
\end{center}
\end{figure}

We can consider only the intraband electron-phonon interaction owing to band gap.
\be
\label{yyy}
\widehat{\mathsf{H}}_{\rm e-ph} =   \frac{\lambda}{\sqrt{N}} \,\sum_\alpha\,
\sum_{\vk,\vq} \hat{\mathsf{c}}^\dag_{\alpha \vk+\vq}  u_\vq \hat{\mathsf{c}}_{\alpha \vk},\quad
\alpha = {\rm c,v}.
\ee
The action corresponding to (\ref{yyy}) is  
\be
\fl
S_{\rm e-ph}  = -\lambda \frac{1}{\sqrt{N}} \int dt  \sum_{\vq,\vk} \bar{\Psi} (\vk+\vq) 
( u^{\rm cl}_\vq \mathrm{I}_4 + u^{\rm q}_\vq  \sigma_x \otimes \mathrm{I}_2 )
\Psi(\vk),
\ee
Now everything boils down to the computation of the following correlation function from which the differential
cross-section can be obtained via (\ref{cross4-1}).
\be
\label{main3-1}
\eqalign{
\fl
&\tilde{C}_{k_{\rm i} \hat{\epsilon}_{\rm i}, k_{\rm f} \hat{\epsilon}_{\rm f}}(t_1,t_2,t_3,t_4)
=\int D[\phi^*,\phi] \,\, e^{+i S_+ /\hbar  -i S_-/\hbar + \delta S_{+-}}  \cr
\fl
 & \times \Big [\vec{J}_{\rm e}( \phi_-^*(\vk_{\rm i},t_1), \phi_-(\vk_{\rm i},t_1)) \cdot \hat{\epsilon}_{\rm i}^* \Big ]  
\Big [\vec{J}_{\rm e}(\phi_-^*(-\vk_{\rm f},t_2), \phi_-(-\vk_{\rm f},t_2)) \cdot \hat{\epsilon}_{\rm f} \Big ]  \cr
\fl
& \times
\Big[ \vec{J}_{\rm e}(\phi_+^*(-\vk_{\rm i},t_3),  \phi_+(-\vk_{\rm i},t_3)) \cdot \hat{\epsilon}_{\rm i} \Big ] \,  
\Big [\vec{J}_{\rm e}(\phi_+^*((\vk_{\rm f}, t_4),\phi_+((\vk_{\rm f}, t_4)) \cdot 
\hat{\epsilon}^*_{\rm f} \Big ],
}
\ee
The correlation function for the  one-phonon Raman scattering [corresponding to (\ref{final}) in equilibrium case with similar notations]
is given by
\be
\label{zz1}
\fl
\eqalign{
\tilde{C}(t_1,t_2,t_3,t_4) & = \lambda^2 \int^\infty_{-\infty} dt_5 \int^\infty_{-\infty} d t_6 \,
M^a_{--}(t_1,t_2,t_5)  [ i D^{ab}_{\vq =0}(t_5,t_6) ] M^b_{++}(t_3,t_4,t_6), \cr
}
\ee
whose Feynman diagram is given in figure \ref{realtime}.
The electron Green's function satisfy the following Dyson equation which includes the pumping term
[ $\widehat{G}^{(0)}_\vk(t-t')$ is the non-interacting Green's function ]
\be
\label{zz2}
\widehat{G}_\vk(t,t') = \widehat{G}^{(0)}_\vk(t-t') + \int^\infty_{-\infty} d t^{\prime \prime}
\widehat{G}^{(0)}_\vk(t-t^{\prime \prime} ) \mathcal{E}(t^{\prime \prime}) \mathrm{I} \otimes \hat{d}
\widehat{G}_\vk(t^{\prime \prime},t').
\ee
(\ref{zz2}) follows from the standard Feynman rules.
The phonon subsystem is also driven into non-equilibrium through the phonon self-energy. Its influence on
ILS can be seen in higer order Feynman diagrams [ see figure \ref{self} (b) ].
To best of our knowledge, the ILS in non-equilibrium condition
has not been presented in the form of the equations (\ref{zz1},\ref{zz2}) before.
The detailed investigation  of (\ref{zz1},\ref{zz2}) is beyond the scope of this paper and  is left for future studies.
\section{Discussions}
\label{summary}

The KS functional integral approach to ILS possesses several advantages  over 
the conventional approach based on Fermi golden rule \cite{basko} even in the lowest order calculations.
First, the finite temperature effect is naturally incorporated as can be seen in (\ref{final1}), which then allows the discussion of
both Stokes and anti-Stokes processes on equal footing.
Second, the energy conservation  for one-phonon process has been  imposed by hand in \cite{basko}, while in our approach 
it is a natural consequence of Feynman rule.

\begin{figure}[!hbt]
\begin{center}
\includegraphics[width = 0.7 \textwidth]{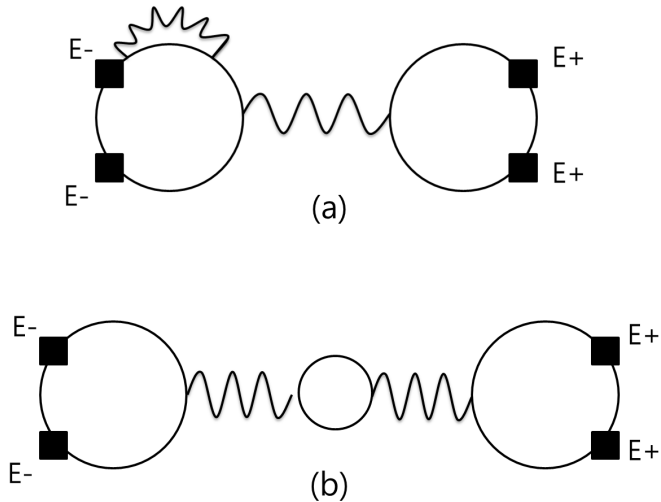}
\caption{Electron (a) and  phonon (b) self-energy contributions to the  Feynman diagram of ILS.}
\label{self}
\end{center}
\end{figure}

\begin{figure}[!hbt]
\begin{center}
\includegraphics[width = 0.7 \textwidth]{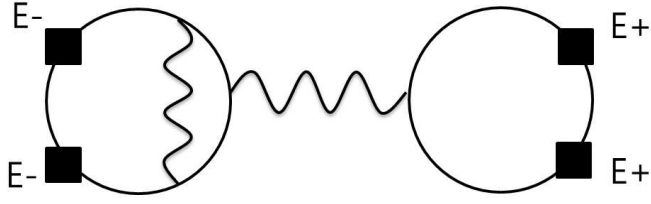}
\caption{A vertex correction contribution to the  Feynman diagram of ILS.}
\label{vertex}
\end{center}
\end{figure}

True advantages of KS functional approach lie in  the study of higher order many-body  effects.
In terms of Feynman diagrams the many-body effects are often described by 
the self-energy corrections and the vertex corrections.
For example, the Feynman diagram of the lowest order electron self-energy correction to ILS is given by 
the diagram (a) of figure \ref{self}, and the diagram (b) describes the the lowest order  phonon self-energy correction.
A lowest order vertex correction to ILS is described in figure \ref{vertex}.
With phonon self-energy corrections included, 
the phonon Green's function in (\ref{final}) can be taken to be fully interacting Green's function, so that 
the phonon frequency shift and the phonon broadening effect  (both stemming from
phonon self-energy) for ILS  can be precisely addressed.

Both self-energy and vertex correction are known to possess the logarithmic corrections \cite{aleiner}, 
which implies that the inclusion of many-body effects are crucial for the proper interpretation of experimental ILS data.
Our formalism provides a natural framework for the investigation of such higher order many effects.
The study of the above many-body effects based on Kramers-Heisenberg formula would be exceedingly difficult,
if not impossible.

In summary, we have expressed the scattering cross section of the  resonant ILS in the form of correlation function
both in equilibrium and non-equilibrium conditions, which then
is recast in the framework of  KS functional integral.
The correlation function in KS functional integral can be computed by the Feynman diagram perturbation theory 
which permits the \textit{systematic} study of many-body effects.

\ack
This work was  supported by the National Research Foundation of Korea (NRF) grant funded by the Korea
government Mest (No. 2012-0008058).
\appendix
\section{The calculations of the photon operator expectation values}
\label{apppendix1}
 The Hamiltonian of photon is given by  ($\alpha$ is the photon polarization)
 \be 
 \label{photonH}
 \sfH_{\rm p} =\sum_{\vk,\alpha} \hbar \omega_\vk ( a_{\vk \alpha}^\dag a_{\vk \alpha} + \frac{1}{2} ) ,
\ee
where $a_{\vk \alpha},a_{\vk \alpha}^\dag$ are the photon annihilation operator and destruction operator, respectively.
The Hamiltonian for the interaction between photon and matter is given by
\be
\label{m-ph}
\sfH_{\sfH_0}'(t) = -\frac{1}{c} \int d \vec{r} \,\vec{J}_{\rm e}(\vec{r},t) \cdot \vec{A}(\vec{r},t),
\ee
where  the operators $ \vec{J}_{\rm e}(\vec{r},t) $ and $ \vec{A}(\vec{r},t)$ have acquired the time dependence  via Heisenberg 
representation with respect to $\sfH_0$.

The second quantized  vector potential operator $ \vec{A}(\vec{r},t)$ of photon is  (in Gaussian unit)
\be
\label{photon}
\vec{A}(\vec{r},t) = \frac{1}{\sqrt{\mathcal{V}}} \sum_{\vk,\alpha} \, \sqrt{\frac{ 2 \pi \hbar c^2}{\omega_\vk}} \,
\Big[ \hat{\epsilon}_{\vk \alpha} a_{\vk \alpha} e^{ i \vk \cdot \vec{r} - i \omega_\vk t} +
 \hat{\epsilon}^*_{\vk \alpha} a^\dagger_{\vk \alpha} e^{- i \vk \cdot \vec{r} + i \omega_\vk t} \Big ], 
\ee
where $\omega_\vk = c \vert \vk \vert$. We recall the standard photon operator commutation relations
\be
\label{commutator}
[ a_{\vk \alpha}, a^\dag_{\vk' \alpha'} ] = \delta_{\vk, \vk'} \delta_{\alpha \alpha'}, \;\;
[ a_{\vk \alpha}, a_{\vk' \alpha'} ]=[ a^\dag_{\vk \alpha}, a^\dag_{\vk' \alpha'} ]=0.
\ee
Employing the commutation relations (\ref{commutator}) it is straightforward to compute the following expectation values 
of photon operators.
\be
\label{expectations}
\eqalign{
& \la \vk_{\rm f},\hat{\epsilon}_{\rm f} \vert a_{\vk c} a^\dag_{\vk' d} \vert \vk_{\rm i},\hat{\epsilon}_{\rm i} \ra=
\delta_{\vk',\vk_{\rm f}} \delta_{d,\hat{\epsilon}_{\rm f}} \delta_{\vk,\vk_{\rm i}} \delta_{c,\hat{\epsilon}_{\rm i}} + \delta_{\vk,\vk'} \delta_{c,d} 
\delta_{\vk_{\rm f},\vk_{\rm i}} \delta_{\hat{\epsilon}_{\rm f},\hat{\epsilon}_{\rm i}}, \cr
&\la \vk_{\rm f}, \hat{\epsilon}_{\rm f} \vert a^\dag_{\vk c} a_{\vk' d} \vert \vk_{\rm i},\hat{\epsilon}_{\rm i} \ra 
= \delta_{\vk',\vk_{\rm i}} \delta_{d,\hat{\epsilon}_{\rm i}} \delta_{\vk,\vk_{\rm f}} \delta_{c,\hat{\epsilon}_{\rm f}}+
\delta_{\vk,\vk'} \delta_{c,d} 
\delta_{\vk_{\rm f},\vk_{\rm i}} \delta_{\hat{\epsilon}_{\rm f},\hat{\epsilon}_{\rm i}} }
\ee
In our case of ILS the initial and the final photon states are different, so that the second terms of 
(\ref{expectations}) vanish.

Now we are ready to compute 
 the operator $\widehat{\mathsf{D}}_{\tilde{k}_{\rm i},\tilde{k}_{\rm f}}$ [ see equations (\ref{d1}) and (\ref{d2}) for
 definitions ]  explicitly.
Plugging (\ref{photon}) into (\ref{m-ph}) and using the result (\ref{expectations})
it is straightforward to obtain ($\omega_{\rm i,f} = c \vert \vk_{\rm i,f} \vert $)
\be
\label{op1}
\eqalign{
\fl
\widehat{\mathsf{D}}_{\tilde{k}_{\rm i},\tilde{k}_{\rm f}}(t_1,t_2)=
\frac{4 \pi \hbar }{\mathcal{V} \sqrt{\omega_{\rm i} \omega_{\rm f}}} \int d\vec{r}_1 d \vec{r}_2 &
\mathtt{T}    \Big[  \{ \vec{J}_{\rm e}(x_1) \cdot \hat{\epsilon}_{\rm i} e^{-i k_{\rm i} \cdot x_1} \} 
\{ \vec{J}_{\rm e}(x_2) \cdot \hat{\epsilon}_{\rm f}^* e^{+i k_{\rm f} \cdot x_2} \} \Big ]
}
\ee
where $x=( \vec{r}, t)$ and $k = (\vk, \omega_\vk)$ are 4-vectors, and
$ k \cdot x = \omega_\vk t -  \vk \cdot \vec{r}$.
Similarly, we can obtain
\be
\label{op2}
\eqalign{
\fl
\widehat{\mathsf{D}}^\dag_{\tilde{k}_{\rm i},\tilde{k}_{\rm f}}(t_3,t_4)= \frac{4 \pi \hbar  }{\mathcal{V} \sqrt{\omega_{\rm i}\omega_{\rm f}
}} \int d\vec{r}_3 d \vec{r}_4 
&\tilde{\mathtt{T} }   \Big[ \{ \vec{J}_{\rm e}(x_3) \cdot \hat{\epsilon}_{\rm i}^* e^{+i k_{\rm i} \cdot x_3} \}
\{ \vec{J}_{\rm e}(x_4) \cdot \hat{\epsilon}_{\rm f} e^{-i k_{\rm f} \cdot x_4} \} 
 \Big ]}
\ee
The  spatial Fourier transform of the electric current operator is defined as
\be
\label{fourier}
J_{\rm e}^a( {\bf k}_j, t_j) = \int d \vec{r}_j \, e^{- i \vk_j \cdot \vec{r}_j} \, J_{\rm e}^a(\vec{r}_j,t_j).
\ee



\begin{thebibliography}{99}
\bibitem{comment1} It can be easily checked that
$ \mathtt{T} (\hat{X}(t) \hat{Y}(t') ) =\pm \mathtt{T} (\hat{Y}(t') \hat{X}(t)  )$
\end{thebibliography}
\end{document}